\documentclass[pra,aps,twocolumn,nofootinbib,superscriptaddress]{revtex4}

\usepackage{graphicx,graphics}
\usepackage{dcolumn}				
\usepackage{amssymb}

\usepackage{hyperref}
\usepackage{bm,pstricks}					

\newcommand{\beqa}{\begin{eqnarray}}
\newcommand{\eeqa}{\end{eqnarray}}
\newcommand{\beq}{\begin{equation}}
\newcommand{\eeq}{\end{equation}}
 
\newcommand{\mean}[1]{\langle #1\rangle}

\begin{document}
\title{Dynamical phases for the evolution of the entanglement between two oscillators coupled to the same environment}

\author{Juan Pablo \surname{Paz}}
\affiliation{Departamento de F\'{\i}sica, FCEyN, UBA, Pabell\'on 1,
Ciudad Universitaria, 1428 Buenos Aires, Argentina}

\author{Augusto J. \surname{Roncaglia}}
\affiliation{Departamento de F\'{\i}sica, FCEyN, UBA, Pabell\'on 1,
Ciudad Universitaria, 1428 Buenos Aires, Argentina}

\begin{abstract}
We study the dynamics of the entanglement between two oscillators that are initially prepared in a general two-mode Gaussian state and evolve while coupled to the same environment. In a previous paper we showed that there are three qualitatively different dynamical phases for the entanglement in the long time limit: sudden death, sudden death and revival and no-sudden death [Paz $\&$ Roncaglia, Phys. Rev. Lett. {\bf 100}, 220401 (2008)]. Here we generalize and extend those results along several directions: We analyze the fate of entanglement for an environment with a general spectral density providing a complete characterization of the evolution for ohmic, sub-ohmic and super-ohmic  environments. We consider two different models for the interaction between the system and the environment (one  where the coupling is through position and another where the coupling is symmetric in position and momentum). Finally, we show that for non-resonant oscillators the final entanglement is independent of the initial state and that it may be non-zero at very low temperatures. 
\end{abstract}

\maketitle
\section{Introduction}

The creation and manipulation of entanglement is an important issue not only because of its fundamental implications but also due to its practical applications. In fact, during the last decade many new discoveries regarding the physics of entanglement were made \cite{Horodecki07}. The development of quantum algorithms and cryptographic schemes were probably the driving forces behind most of the research on entanglement. In fact, nowadays entanglement is regarded not only as a peculiar feature of quantum systems but also as a physical resource. Entanglement manipulation was studied first for finite dimensional systems but later continuous variable systems \cite{Braunstein05} were considered. In fact, several experiments showed the successful implementation of quantum teleportation \cite{Furu98} and cryptographic protocols \cite{Grosshans03} for such systems. In this context, it is important to take into account the effects induced by the interaction between a composite (eventually entangled) quantum system and its surrounding environment. Indeed, decoherence in some cases can be devastating: Thus, due to the interaction with the environment, entanglement within a composite system can disappear in a finite time. This phenomenon, that was first discussed and analyzed for systems made out of qubits \cite{Diosi03,Yu04,Almeida07} became known as ``sudden death'' of entanglement (SD). But the fate of entanglement for a quantum open system is not at all evident and some surprising results were also obtained: For example, it was shown that under certain conditions the environment can act as a quantum channel thorough which entanglement can be created \cite{Braun02}. In this case, even if the initial state of the system is separable, the final state could be entangled. A large number of recent papers study this and other issues that characterize the dynamics of entanglement in systems of qubits interacting with common or independent environments \cite{Braun02,Kim02-2,Benatti03,Oh06,Anastopoulos06,Bellomo07,Bellomo08}.
On the other hand, continuous variable systems were also investigated and similar results emerged. For example, the degradation of entanglement for harmonic systems interacting with bosonic reservoirs was analyzed \cite{Paris02,Serafini04,Olivares07}. Also, the fate of initially entangled states (two-mode squeezed states) interacting with a common bath was studied using different approximations \cite{Dodd04,Prauzner04,Benatti06-2}. Among the interesting results that emerged from those works it is worth mentioning that in \cite{Prauzner04} a condition for the existence of sudden death  was deduced under the RW-Markovian assumption. More recently, the non-Markovian regime was also 
analyzed \cite{Liu07,An07,Horhammer07} and a zoo with different long time behavior emerged. Thus, it was realized in \cite{Liu07} that non-Markovian effects modify the condition obtained in \cite{Prauzner04} for the existence of SD.

More recently, \cite{Paz08}, we provided a unified picture of the different qualitative dynamics of entanglement for general gaussian states in non-Markovian environments. There, we showed that the asymptotic dynamics of entanglement can be described by three possible phases: SD (sudden death), SDR (sudden death and revival) and NSD (no sudden death).
The existence of an exact master equation for quantum Brownian motion enabled us to obtain analytical expressions for the asymptotic entanglement and for the  boundaries between the phases. In the present paper we complete and generalize the  ideas presented in \cite{Paz08}. Here, we consider two different models for the coupling between the system and the environment: We analyze not only the case where the coupling is bilinear both in the position of the system and the environment but also we solve the case where the coupling between the system and the environment is symmetric between position and momentum (this is technically equivalent to the RWA). We analyze in detail how the phase diagram changes depending on the coupling to the environment as well as on the environmental spectral density. For both models we study the entanglement for ohmic, sub-ohmic and super-ohmic spectral densities. Finally, we study the entanglement between non-resonant oscillators where a new master equation is derived. In such case, we show that although non-resonant effect tend to eliminate entanglement, it is possible to have resilient entanglement at sufficiently low temperatures. 

The paper is organized as follows. In Section \ref{sec:model} we review the basic technical tool we will use in our analysis: the master equation. In Section \ref{sec:EvolEntan} we show how to use this equation to analyze the evolution of the entanglement for general Gaussian states. In Section \ref{sec:phasediagram} we present a detailed analysis of all qualitatively different evolutions (dynamical phases) of entanglement. 
In Section \ref{sec:offres} we study the evolution of entanglement for  non-resonant oscillators. In Section \ref{sec:conc} we summarize and conclude. 

\section{Two exactly solvable models}
\label{sec:model}

We will study the evolution of the entanglement between two harmonic oscillators with coordinates $x_1$ and $x_2$ (they constitute our system) which are coupled with a bosonic environment. We will analyze two different models: First we will assume that the coupling between the system and the environment is bilinear in their position \cite{FeyVer63,CalLeg83,Grabert88,HuPazZha92}. Then, we will analyze the case where the coupling is symmetric in position and momentum. In both cases we will use an exact master equation to describe the evolution of the reduced density matrix of the system. In what follows we will briefly describe the two models and their solution. 

\subsection{Quantum Brownian motion with position coupling}

The total Hamiltonian for the universe formed by the system and the environment is 
$H=H_S+H_{int}+H_{env}$ where  
\beqa
H_S&=&\frac{p_1^2+p_2^2}{2m}+\frac{m}{2}(\omega_1^2 x_1^2+\omega_2^2 x_2^2)+mc_{12} 
x_1 x_2, \nonumber \label{eq:H_ent}  \\
H_{env}&=&\sum_{n=1}^{N}(\frac{\pi_n^2}{2m_n}+\frac{m_n}{2} w_n^2 q_n^2),\\
H_{int}&=&(x_1+x_2)\sum_{n=1}^{N}c_n q_n.\nonumber\label{eq:positioncoupling}
\eeqa
It is convenient to use coordinates $x_\pm=(x_1\pm x_2)/\sqrt{2}$ since $x_+$ couples to the environment. The Hamiltonian $H_S$ is 
\beq
H_S=\frac{(p_+^2+p_-^2)}{2m} + \frac{m}{2}(\omega_-^2 x_-^2 +\omega_+^2 x_+^2)+
m c_{+-} x_+x_-, \nonumber
\eeq
where the frequencies of the $x_\pm$ oscillators are $\omega_{\pm }^2=(\omega_1^2+\omega_2^2)/2\pm c_{12}$ and the coupling constant between them is $c_{+- }=(\omega_1^2-\omega_2^2)/2$. Below we will analytically solve a special but very important case: We consider the two oscillators to be resonant, i.e. 
we take $\omega_1=\omega_2$ (in this case the $x_\pm$ oscillators are decoupled, as $c_{+-}=0$). 

This model (known as Quantum Brownian Motion) can be exactly solved  \cite{HuPazZha92}. Thus only two parameters are necessary to characterize the effect of the environment on the system. The first one is the initial state of the environment (assumed to be thermal, with initial temperature $T$). The second one is the spectral density of the environment, which is a function of the frequency defined as $J(\omega)=\sum_n c_n^2\delta(\omega-w_n)/2m_n  w_n$. One can show that the reduced density matrix $\rho$, obtained from the state of the universe by tracing out the environmental oscillators, obeys an exact master equation which is written as \cite{HuPazZha92,Chou07}:
\beqa
\dot\rho&=&-i[H_R,\rho]-i\gamma(t)[x_+,\{p_+,\rho\}]-\nonumber\\
&-&D(t)[x_+,[x_+,\rho]]-f(t)[x_+,[p_+,\rho]].\label{eq:mastereq}
\eeqa
Here, the renormalized Hamiltonian is 
\beq 
H_R=H_S+{m\over 2}\delta\omega^2(t)x_+^2.\nonumber
\eeq 
The coefficients $\delta\omega^2(t)$, $\gamma(t)$, $D(t)$ and $f(t)$ depend on the 
spectral density of the environment ($D(t)$ and $f(t)$ also depend on the initial temperature $T$). The explicit form of these coefficients is rather cumbersome  
and was studied in detail elsewhere \cite{HuPazZha92,Fleming07}. Some results on the behavior of the coefficients for typical environmental spectral densities will be described below. In particular we will consider the family of spectral densities of the form:
\beq
J(\omega)={2\over \pi}m\gamma_0\omega \Big({\omega\over\Lambda}\Big)^{n-1} \theta(\Lambda-\omega),
\label{eq:spectral}
\eeq
where $\Lambda$ is the cutoff frequency and $\gamma_0$ is a coupling constant. 
Depending on the value of $n$, the spectral densities are known as: ohmic $(n=1)$, 
sub-ohmic $(n<1)$ and super-ohmic $(n>1)$.

To study analytically the long-time regime,
we just need to assume (as it is the case for realistic environments) that
the coefficients of the master equation approach asymptotic values after a temperature-dependent time. The time dependent frequencies $\Omega^2_{1,2}(t)=\omega^2_{1,2}+\delta\omega^2(t)/2$ approach cutoff independent
values only if the bare frequencies $\omega_{1,2}$ have an appropriate dependence on the cutoff. The coupling constant $c_{1,2}$ must also be renormalized in the same way so that the time 
dependent coupling $C_{12}(t)=c_{12}+\delta\omega^2(t)/2$ approaches a finite cutoff independent value. The behavior of the diffusion coefficients $D(t)$ and $f(t)$ is more 
complicated and depend on the initial temperature. 
A word on notation: upper case letters will be used to denote renormalized quantities. 
The time label will be omitted when referring to the asymptotic value of 
the corresponding function (i.e., $\Omega_{1,2}$ denotes the asymptotic value 
of the renormalized frequency of the oscillators, etc). 

The master equation is a powerfull tool to understand the qualitative behavior of the system. For this purpose, it is convenient to use it to obtain simple evolution equations for the second moments of $x_{\pm}$ and $p_{\pm}$. Thus, it is simple to show that the second moments of $x_+$ and $p_+$, satisfy the following equations:
\beq
{d\over{dt}}
\left({\langle p_+^2 \rangle\over{2m}}\right)+ 
\frac{m}{2} \Omega^2(t)
{d\over{dt}}      
\langle x_+^2 \rangle =
-\frac{2\gamma(t)}{m}\langle p_+^2\rangle+{D(t)\over{m}}, 
\eeq
\beq
{1\over 2}\frac{d^2\langle x_+^2\rangle}{dt^2}
+\gamma(t)\frac{d\langle x_+^2\rangle}{dt}
+\Omega^2(t)\langle x_+^2\rangle = 
\frac{\langle p_+^2\rangle}{m^2}-\frac{f(t)}{m}. \nonumber\label{eq:momenta}
\eeq
Where $\Omega(t)$ is the renormalized frequency of the $x_+$ oscillator.
In turn, the evolution equations for the second moments of $x_-$ and $p_-$ are simply the ones of a free oscillator (i.e., can be obtained from the above ones by considering vanishing values for all the coefficients of the master equation). 

From the above equations the interpretation of the coefficients appearing in the master equation is transparent: $\gamma(t)$ is responsible for relaxation since it induces the decay of energy, $D(t)$ is a normal diffusive term which induces heating increasing the momentum dispersion. In turn, $f(t)$ the so-called anomalous diffusion coefficient is responsible for the squeezing of the asymptotic state (see below) or, in other words, of a violation of the equipartition principle: Thus, in the stationary state (which is reached only if the environment is such that the coefficients approach constant asymptotic values) eq. (\ref{eq:momenta}) implies that the expectation value of kinetic and potential energy differ by a factor which is proportional to $f(t)$. The role of this term, will be very important in our analysis below. 

Our analysis will be based on the use of the above equations to study the long time regime for cases where the environment is such that the coefficients of the master equation do approach a constant asymptotic value. Thus, it will be useful to write down the explicit asymptotic values of the dispersions 
$\Delta^2 x_+=\langle x_+^2\rangle$ and $\Delta^2 p_+=\langle p^2_+\rangle$. From the above equations we find that
\beq
\Delta p_+= \sqrt{{D \over 2 \gamma}}, \quad \Omega \Delta x_+= \sqrt{{D \over 2 m^2\gamma} - 
{f \over m}}, \label{eq:Dandf}
\eeq
and $\mean{\{x_+,p_+\}}=0$. It is worth noticing that depending on the sign of the asymptotic value of $f$ 
the nature of the relation between the variances, or squeezing, may change quite dramatically. 
The sign of $f$ indicates what observable is being effectively localized.
In fact, if the coefficient $f$ is positive the asymptotic
state is localized in position (i.e., the equilibrium state is squeezed along position), which is a feature
of low temperatures.

\subsection{Quantum Brownian motion with coupling symmetric in position and momentum}

We will also consider another exactly solvable model which is very similar to the above one. The only difference is that the system and the environment are coupled through different observables. The interaction Hamiltonian between the two resonant oscillators and the environment is
\beq
\tilde H_{int}=(x_1+x_2)\sum_{n=1}^{N} c_n q_n+
\left(\frac{p_1+p_2}{m\omega}\right)
\sum_{n=1}^{N} {\tilde c_n\over m_n w_n} \pi_n.\nonumber
\eeq
In the case $c_n=\tilde c_n$ the total interaction can be rewritten in terms of creation and anihillation operators of the $x_+$ oscillator (denoted $a$ and $a^\dagger$) and the environmental ones (denoted as $b_n$ and $b_n^\dagger$). Thus,
\beq
\tilde H_{int}=\sum_{n=1}^{N} \frac{c_n 2\sqrt{2}}{\sqrt{m m_n \omega w_n}} 
(a b_n^\dagger+a^\dagger b_n) .\nonumber
\label{eq:HRWA}
\eeq
This is the same type of interaction that one obtains by making the so-called rotating wave approximation (RWA) for the model with Hamiltonian (\ref{eq:positioncoupling}). It is worth pointing out that we will discuss this as a separate model with its own exact solution (and not necessarily as an approximation to the previous one). Here, the system interacts with the environment both through position and momentum. As in the previous case, interactions within the system are induced through the environment. Such  interactions generate a renormalization of the system's parameters. To be able to properly renormalize all the parameters in the Hamiltonian of the system we should include the most general type of interactions in such Hamiltonian. In this case, it includes not only coupling between the oscillators coordinates but also momentum coupling. Thus, the Hamiltonian of the system is
\beqa
\tilde H_S&=&\frac{p_1^2+p_2^2}{2m}+\frac{m}{2}\omega^2(x_1^2+x_2^2)+\nonumber\\
&+&mc_{12} x_1 x_2 \nonumber 
+ \frac{\tilde c_{12}}{m \omega^2} p_1 p_2. \nonumber 
\eeqa
In the resonant case we are considering here, this Hamiltonian is simply written in terms of coordinates $x_\pm$ as the sum of two decoupled oscillators with frequencies $\omega_{\pm }^2=\omega^2(1\pm c_{12}/\omega^2) (1\pm\tilde c_{12}/\omega^2)$ and masses 
$m_\pm=m/(1\pm \tilde c_{12}/\omega^2)$.

In this case, it is possible to obtain an exact master equation for the reduced density matrix $\rho$. For the zero temperature case the exact master equation was obtained by An et al \cite{An07-2}. Their result can be generalized to finite temperature (details of the derivation will be presented elsewhere) and reads:
\beqa
\dot\rho&=&-i[\tilde H_R,\rho]-i\tilde \gamma(t)\Big([x_+,\{p_+,\rho\}]-[p_+,\{x_+,\rho\}]\Big)\nonumber\\
&-&\tilde D(t)\Big([x_+,[x_+,\rho]]+ {1\over m_+^2 \omega_+^2} [p_+,[p_+,\rho]]\Big).\label{eq:mastereq2}
\eeqa
Here, the renormalized Hamiltonian $\tilde H_R$ is
 \beq
 \tilde H_R=\tilde H_S+ \delta\tilde\Omega^2(t)\Big(\frac{1}{2} {p_+^2\over{m_+\omega_+^2}}+\frac{1}{2} m_+ x_+^2\Big).
 \eeq

The main features of this master equation are simple to understand: Not surprisingly, this equation looks as the symmetrized version of (\ref{eq:mastereq}). Thus, the damping coefficient $\tilde\gamma (t)$ appears multiplying a term that is symmetric under canonical interchange of position and momentum. This is also the case for the normal diffusive term (proportional to $\tilde D(t)$). The absence of the anomalous diffusion is precisely an expected consequence of the same symmetry since this term is anti--symmetric   in (\ref{eq:mastereq}). Renormalization is also symmetric since this type of coupling induces renormalization not only on the oscillator frequency but also on its mass.
In fact, renormalized frequencies and masses of each oscillator can be defined as:	
$\Omega_i(t)=\omega(1+\delta\tilde\Omega^2(t)/2\omega^2)$,
$M_i(t)=m/(1+\delta\tilde\Omega^2(t)/2\omega^2)$. In turn, renormalized coupling constants are: $C_{12}(t)=c_{12}+\delta\tilde\Omega^2(t)/2$, 
$\tilde C_{12}(t)=\tilde c_{12}+\delta\tilde\Omega^2(t)/2$.

From the master equation we can obtain equations of motion for the 
second moments of the $x_+$ oscillator:
\begin{widetext}
\beqa
{d\over dt} \mean{p_+^2}&=& 
      -M(t) \Omega^2(t)\mean{\{x_+,p_+\}}-4 \tilde\gamma(t) \mean{p_+^2}+ 2 \tilde {D}(t),\nonumber\\
{d\over dt}\mean{x_+^2}&=&{1\over M(t)}\mean{\{x_+,p_+\}}-4 \tilde\gamma(t) \mean{x_+^2}+{2\over M(t)^2\Omega^2(t) } \tilde{D}(t),\\
{d\over dt}\mean{\{x_+,p_+\}}&=&2{\mean{p_+^2}\over M(t)}-2M(t)\Omega^2(t)\mean{x_+^2}-4 \tilde\gamma(t) \mean{\{x_+,p_+\}}.\nonumber
\eeqa
\end{widetext}
where $M(t)=m/(1+(\delta\tilde\Omega^2(t)+\tilde c_{12})/\omega^2)$ and $\Omega(t)=\omega(1+(\delta\tilde\Omega^2(t)+c_{12})/\omega^2)$ are the mass and the frequency of the oscillator $x_+$. 
The role of each term is transparent: $\tilde\gamma(t)$ is an effective damping rate inducing decay towards the ground state while $\tilde D(t)$ is a symmetrized diffusion constant inducing the spread of the state both in position and momentum. Assuming these coefficients approach constant asymptotic values we can easily derive the long time values of position and momentum dispersions to be given by 
\beq
\Delta p_+=M\Omega\Delta x_+= \sqrt{{\tilde{D} \over 2 \tilde\gamma}}; \label{eq:RWADxDp}
\eeq
and $\mean{\{x_+,p_+ \}}=0$. 

Contrary to what happened in the non-symmetric case, governed by the master equation (\ref{eq:mastereq}), the asymptotic state satisfies the equipartition principle since the expectation values of kinetic and potential energies are identical. Analogously, as will be mentioned below, the asymptotic state of the $x_+$ oscillator is not squeezed.

\section{Evolution of entanglement for Gaussian states}
\label{sec:EvolEntan}

We will assume that the initial state of the system is Gaussian. As the complete evolution is linear, the Gaussian nature of the state will be preserved for all times. This enables us to analytically compute the entanglement between the two oscillators in the following way: Entanglement for Gaussian states is entirely determined by the properties of the covariance 
matrix defined as 
\beq
V_{ij}(t)=\mean{\{r_i,r_j\}}/2-\mean{r_i}\mean{r_j},\nonumber
\eeq
where $i,j=1,\ldots,4$ and $\vec r=(x_1,p_1,x_2,p_2)$. In fact, a good measure of entanglement for such states is the so-called logarithmic negativity $E_{\mathcal N}$ \cite{Vidal02,PhdEisert} which can be computed as \cite{Vidal02,PhdEisert,Adesso04}: 
\beq
E_{\mathcal N}=\max\{0,-\ln(2\nu_{\min})\},
\eeq
where $\nu_{\rm min}$ is the smallest symplectic eigenvalue of the partially transposed covariance matrix. There are known expressions for $E_{\mathcal N}$ for particularly relevant Gaussian states which will be used as initial conditions in our study. For this reason it is useful to mention them here: For the two-mode squeezed state, obtained 
from the vacuum by acting with the operator $\exp(-r(a_1^\dagger a_2^\dagger- a_1 a_2))$,  
we have $E_{\mathcal N}=2|r|$. For this state the dispersions satisfy the minimum 
uncertainty condition $\delta x_+ \delta p_+ =\delta x_- \delta p_- =1/2$. The squeezing 
factor determines the ratio between variances since 
$m\Omega\delta x_+ /\delta p_+ =\delta p_- /(m\Omega\delta x_-) =\exp(2r)$. 
As $r\rightarrow\infty$ the state becomes localized 
in the $p_+$ and $x_-$ variables approaching an ideal EPR state \cite{EPR}. 

Now we consider a general initial gaussian state of the two oscillators.
From the appropriate master equation (\ref{eq:mastereq}) or (\ref{eq:mastereq2}) 
we showed how to obtain equations for the covariances. These equations split into two 
blocks of $2\times 2$. The evolution of the first block 
formed with the second moments of $x_-$ and $p_-$ corresponds to a free oscillator with frequency $\omega_-$,
which can always be expressed in terms of two dispersions $\delta x_-$ and $\delta p_-$. 
The evolution equations for the second block, formed with the second moments of $x_+$ and $p_+$, 
were discussed above and yield equilibrium values $\Delta x_+$ and $\Delta p_+$. It can be also 
easily proved that the off-diagonal block, containing the correlations between 
the oscillators $(x_+,x_-)$, vanishes in the asymptotic regime. 
These simple observations are almost all we need to fully analyze the evolution of the entanglement between initial Gaussian states. Thus, using the diagonal block form of the covariance matrix in the $(x_+, x_-)$ bases (and changing basis to obtain covariances of the original $x_{1,2}$ oscillators) it is simple to find the smallest symplectic eigenvalue of such matrix and compute the logarithmic negativity. The result is: 
\beq
E_{\mathcal N}(t)\rightarrow\max\{0,E(t)\}, 
\eeq
where the function $E(t)$ is defined as
\beq
E(t)= \tilde E_{\mathcal N}+{\Delta E_{\mathcal N}} G(t). \label{eq:Eoft}
\eeq
Here $G(t)$ is an oscillatory function with period $\pi/\omega_-$ that takes values in the interval $\{ -1,+1 \}$. Its explicit form will be given below.  The mean value $\tilde E_{\mathcal N}$ and the amplitude $\Delta E_{\mathcal N}$ that characterize the oscillations of $E(t)$ are simply written as
\beqa
\tilde E_{\mathcal N}&=&\max\{|r|,|r_{crit}|\} -S_{crit},\label{eq:ent}
\\
\Delta E_{\mathcal N}&=&\min\{|r|,|r_{crit}|\}.
\label{eq:deltaE}
\eeqa
In the above equations $r$ is the initial squeezing factor defined as 
\beq
r={1 \over 2}\ln\left[m_- \omega_- {\delta x_- \over \delta p_-}\right],\nonumber
\eeq 
and $r_{crit}$ is related to the squeezing factor of the equilibrium state for the  $x_+$-oscillator
\beq
r_{crit}={1 \over 2}\ln\left[m_- \omega_- {\Delta x_+ \over \Delta p_+}\right]. 
\label{eq:rcrit}
\eeq

Finally, $S_{crit}$ is defined as
\beq
S_{crit}={1\over 2} \ln[4 \Delta x_+ \Delta p_+ \delta x_- \delta p_-], 
\label{eq:Scrit}
\eeq
and turns out to be simply related with the entropy of the asymptotic state. Thus, 
the von Neumann entropy of the final state ($S_v$) is $S_v =f(\sigma_+)+f(\sigma_-)$ where $f(\sigma)=(\sigma +{1 \over 2})\ln(\sigma +{1 \over 2})-(\sigma -{1 \over 2})\ln(\sigma -{1 \over 2})$, with 
$\sigma_+=\Delta x_+ \Delta p_+$ and $\sigma_-=\delta x_- \delta p_-$.
It is worth mentioning that in all the above formulae the dispersions $\Delta x_+$ and $\Delta p_+$ are the asymptotic 
values of the dispersions along position and momentum that depend upon the temperature and the 
type of coupling to the environment (and that, for the models analyzed above are given by 
eqs. (\ref{eq:Dandf}) and (\ref{eq:RWADxDp})). For completeness we give the explicit formula for $G(t)$ which is, indeed, not very illuminating:
\begin{widetext}
\beqa
 &&\Delta E_{\mathcal N} G(t)=\max\{|r|,|r_{crit}|\}+\frac{1}{2}\ln\Big[\cosh[2(r-r_{crit})] \cos^2(\omega_- t) +\cosh[2(r+r_{crit})] \sin^2(\omega_- t) 	 \nonumber \\
&-&\sqrt{2\big(\sinh^2(2r)+\sinh^2(2r_{crit})\big) \sin^2(\omega_- t)\cos^2(\omega_- t)+
\sinh^2[2(r_{crit}-r)]\cos^4(\omega_- t)+\sinh^2[2(r_{crit}+r)]\sin^4(\omega_- t)}\Big].	\nonumber
\eeqa
\end{widetext}

These simple results will enable us to draw the following conclusions about
the dynamics of entanglement for long times. 1) Evolution of $E_\mathcal{N}$ is fully characterized by $r_{crit}$ and $S_{crit}$. 2) Only three qualitatively different scenarios emerge. First, entanglement may persist for arbitrary long times. This phase, which we call ``NSD'' (for no-sudden death), is realized when the initial state is such that $\tilde E_{\mathcal N}-\Delta E_{\mathcal N}>0$, which translates into $||r|-|r_{crit}||>S_{crit}$. 
Then, there is a phase where entanglement undergoes an infinite sequence of events of ``sudden death'' and  ``sudden revival'' \cite{Yonac06,Yonac07}. This occurs if the initial state is such that $|E_c|\le r\le -E_c+2|r_{crit}|$,  where the quantity $E_c$ is defined as
\beq
E_c\equiv |r_{crit}|-S_{crit}.\label{eq:Ec}
\eeq 
We denote this phase as ``SDR'' (for sudden death and revival). Finally, a third  
phase characterized by a final event of ``sudden death'' of entanglement may 
be realized if  $|r|\le -E_c$. This phase is simply denoted as ``SD'' (for sudden death). In what follows we will analyze these phases for different spectral densities and coupling strength between the oscillators and the environment.

Some further physical insight about the origin of the entanglement can 
be obtain by rewriting the eq. (\ref{eq:Eoft}) as
\beqa
E(t)&=&|r_{crit}|-S_{crit}+|r| G(t), \ \ \ \ {\rm if}\ |r|\leq |r_{crit}|,\nonumber\\
E(t)&=&|r|-S_{crit}+|r_{crit}| G(t), \ \ \ \ {\rm if}\ |r| > |r_{crit}|.\nonumber
\eeqa
In this way it is clear that for initial values $|r|\le |r_{crit}|$ the environment supplies the resource to generate entanglement. In particular, when $|r_{crit}|-S_{crit}\ge 2|r|$, then the entanglement in the final state is larger than the quantum resource (squeezing) available in the initial state. In other cases the environment does not act as the supplier but simply degrades the quantum 
resource which is already present in the initial state (either in the form of squeezing or entanglement). Below, we will analyze this further by using a very convenient tool: a phase diagram where the fate of entanglement can be graphically depicted for all initial states.

\section{Evolution of entanglement}
\label{sec:phasediagram}

\subsection{Phase diagrams for entanglement dynamics}

\begin{figure}[htb!]
\vspace{0.8cm}
\includegraphics[width=8.7cm]{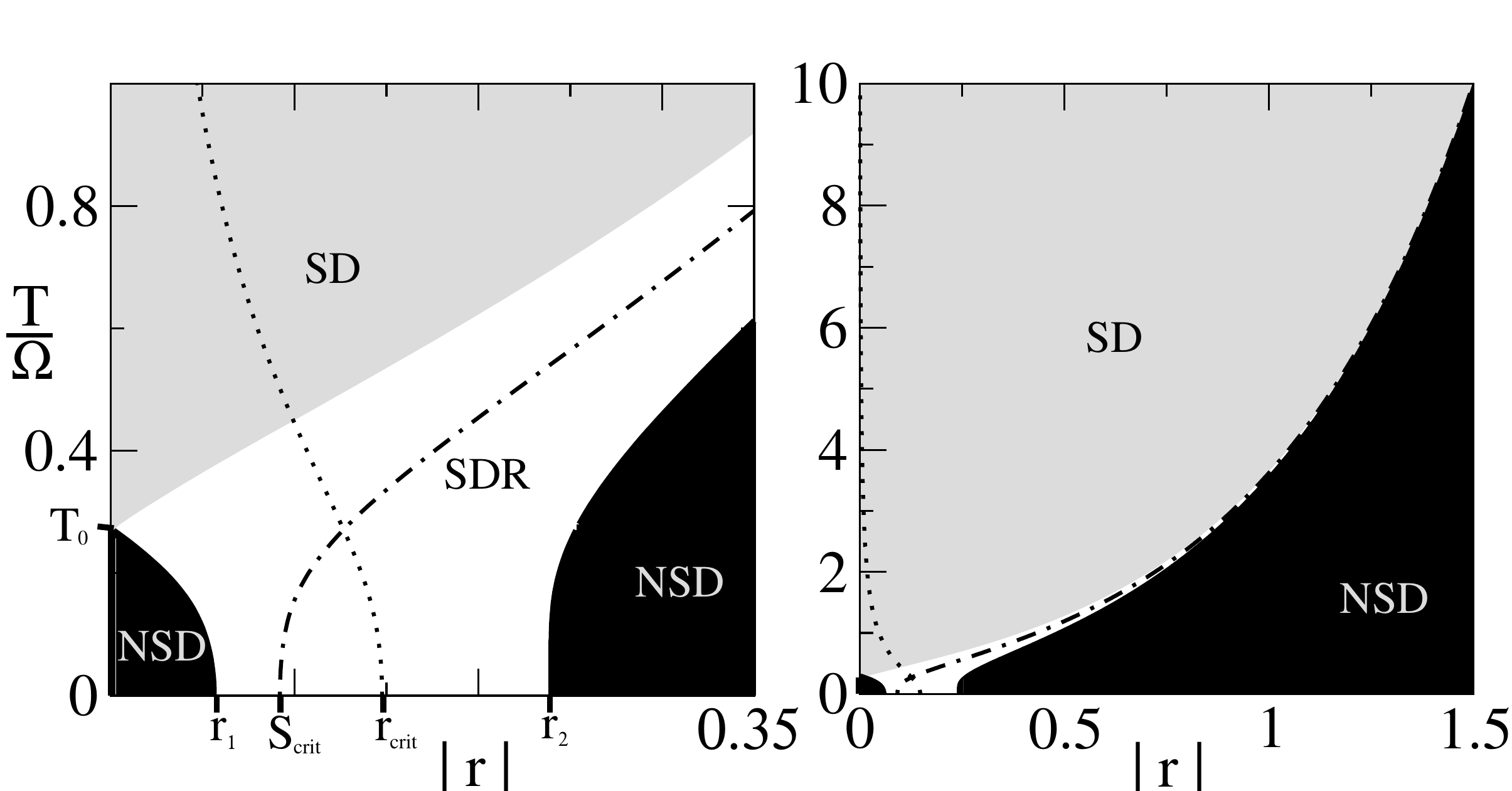}
\caption{Phase diagram for ohmic environment ($\Omega=1$, $m=1$, $\gamma_0=0.1$, 
$\Lambda=20$, $C_{12}=0$, $\delta x_-\delta p_-=1/2$). 
The sudden death (SD), no-sudden death (NSD) and sudden death and revival (SDR) phases 
describe the three different qualitative long time behaviors for the entanglement 
between two oscillators interacting with the same environment. The SDR phase is centered 
about the dashed line $S_{crit}$ and has a width given by the dotted line $|r_{crit}|$. 
This is the case for temperatures above $T_0$, the one for which $S_{crit}=|r_{crit}|$. 
Below this temperature the role of $S_{crit}$ and $|r_{crit}|$ are interchanged. 
SDR separates the SD and NSD phases. The low temperature NSD island is due to non--Markovian and 
non--perturbative effects. $\tilde E_{\mathcal N}$  in the NSD phase is the distance
to the dashed line for $|r| > |r_{crit}|$, and the distance between the  dashed and dotted lines for $|r|\leq |r_{crit}|$.
} 
\label{fig:phases}
\end{figure}

Here we will introduce a convenient tool to study the different dynamical phases of entanglement. In fact, depending on the properties of the environment (initial temperature, damping rate, etc) a given 
initial state (parameterized by the squeezing $r$ and by the product of initial dispersions $\delta x_-\delta p_-$) will belong to one of the three phases: SD, NSD or SDR. For fixed values of $\gamma$ and $\delta x_-\delta p_-$ we 
can always draw a phase diagram like the one displayed in Fig. \ref{fig:phases}. To obtain it we need to analyze the 
temperature dependence of the asymptotic dispersions to obtain both $|r_{crit}|$ and $S_{crit}$ as a function of the
temperature. In the phase diagram, the areas corresponding to each of the three phases are displayed. As a reference,
we also include two curves that show the temperature dependence of $S_{crit}$ and $|r_{crit}|$ (dashed and dotted lines respectively). The actual diagram shown in Fig. \ref{fig:phases} corresponds to a particular case: an environment with 
ohmic spectral density coupled to the system through position with $C_{12}=0$ (then, $\omega_-=\Omega$ and $m_-=m$). 
We also assumed a pure initial state  with $\delta x_-\delta p_-=1/2$.  
In the following sections we will see that other spectral densities will give 
rise to slightly different features in the phase diagram but its topology will remain unaffected. 
Changes in the initial state (i.e., considering mixed states with $\delta x_-\delta p_-> 1/2$) 
can also be simply understood and will be discussed below. 

The phase diagram describes all dynamical information of the asymptotic evolution of entanglement for the case of position coupling to the environment (see below for symmetric coupling). Some important features of the phase diagram are worth mentioning. In particular, we would like to focus first on the NSD phase present at low temperatures. Its origin is purely non-Markovian and non--perturbative. Its area shrinks as the damping rate decreases. The states in this phase are the ones for which the final entanglement may be larger than the squeezing invested in the initial state. For such states the entanglement mostly comes from the squeezing available in the environment. This is particularly clear for the case of coherent states, that can become entangled below the critical temperature $T_0$ (see below). 

To understand the nature of entanglement in this region of the phase diagram it is useful to focus first on the properties of such diagram along with of its axis. The zero temperature line (i.e., the horizontal axis) contains states in the NSD phase for small and large squeezings. Thus, the NSD phase is realized at zero temperature when the initial squeezing $r$ is either 
$|r|\ge r_2$ or $|r|\le r_1$, where
\beqa
r_1&=&{1\over 2} \ln\left[{1\over{2m\Omega\Delta x_+^2(T=0)}}\right],\label{r1}\\
r_2&=&{1\over 2} \ln\left[{2\Delta p_+^2(T=0)\over{m\Omega}}\right],\label{r2}
\eeqa
see Fig. \ref{fig:phases}.
For the range of squeezings between $r_1$ and $r_2$ (the region centered about $|r_{crit}|$) the states belong to the SDR phase. This implies that pure initial states (at $T=0$) will never experience a sudden death. They will never belong to the SD phase. 

It is interesting to notice that for $T=0$ the asymptotic state of the $x_+$ oscillator is squeezed in position (i.e., $m\Omega\Delta x_+(T=0)<\Delta p_+(T=0)$) and that it also has non-vanishing entropy (i.e., $\Delta  x_+(T=0)\Delta p_+(T=0)>1/2$). In the following section we will present analytic expressions for $\Delta x_+$ and $\Delta p_+$ in the case of the ohmic environment. Here it is sufficient to mention that the squeezing in position is a consequence of the fact that, being the interaction with the environment through the position observable, the asymptotic state tends to localize more along 
position than along momentum. The squeezing $r_1$ ($r_2$) is precisely the ratio between the asymptotic position (momentum) dispersion and the one corresponding to the vacuum: a non-vanishing value of $r_1$ means that the asymptotic 
state of the oscillator $x_+$ has a dispersion along position that is smaller than the vacuum. Therefore, the states
that belong to the low temperature NSD island are the ones for which the state of the oscillator $x_+$ is narrower
in position than the vacuum dispersion which is given by $1/2m\Omega$.
As the temperature of the environment increases, the asymptotic value of the position dispersion $\Delta x_+$ also grows. Therefore, the NSD phase shrinks and completely disappears above the critical temperature $T_0$, which is precisely the one for which the position dispersion becomes identical to the one corresponding to the vacuum, i.e.
\beq
T_0\ {\rm such\ that\ }\  \Delta x_+(T=T_0)={1\over\sqrt{2m\Omega}}.\label{T0}
\eeq

The vertical axis of the phase diagram it is also worth analyzing since it describes the fate of coherent states (for which $r=0$). As mentioned above, for temperatures lower than $T_0$ such states end up entangled due to the interaction with the environment. But for temperatures larger than $T_0$ such states always experience an event of sudden death of entanglement (the states belong to the SD phase). 

The high temperature region of the diagram is rather different than the low temperature one. Thus, for high temperatures we have $E_c<0$ (which implies that coherent states do not get entangled) and also $r_{crit}\ll S_{crit}$ (which implies that the region covered by the SDR phase becomes relatively narrower). Hence, initial states with large squeezing ($|r|>\ln(2\Delta x_+\Delta p_+)/2=S_{crit}$) retain some of their entanglement while those with squeezing factors smaller than the critical value $S_{crit}$ suffer from sudden death. However, our analysis shows that the boundary between SD and NSD phases is rather subtle: for any finite temperature  the two phases are separated by a very narrow portion of SDR phase (in this phase there are oscillations of the entanglement whose amplitude, $|r_{crit}|$, depends on the temperature in a way which is different for different spectral densities, as will be discussed below). In any case, these oscillations are, indeed, yet another interesting non-Markovian effect identified by our analysis. 

A final comment on the phase diagram: The NSD phase is 
characterized by a non-vanishing asymptotic entanglement that can be quantified in a straightforward way from the 
phase diagram itself. The average value of the logarithmic negativity is simply the distance to 
the dashed line (which signals the midpoint of the SDR phase) or just the distance between the 
dashed and dotted lines for $|r|\leq|r_{crit}|$.

There is an important qualitative difference between the cases of position coupling and symmetric coupling. In the later case, the symmetry implies that the asymptotic state of the $x_+$ oscillator is not squeezed as $r_{crit}=0$. The asymptotic entanglement is $E_{\mathcal N}(t)=\max\{0,E(t)\}$, where:
\beq
E(t)=|r|-\frac{1}{2}\ln[4\Delta x_+ \Delta p_+ \delta x_- \delta p_-].
\label{eq:EntRWA}
\eeq

\begin{figure}[htb!]
\vspace{0.8cm}
\includegraphics[width=4.3cm]{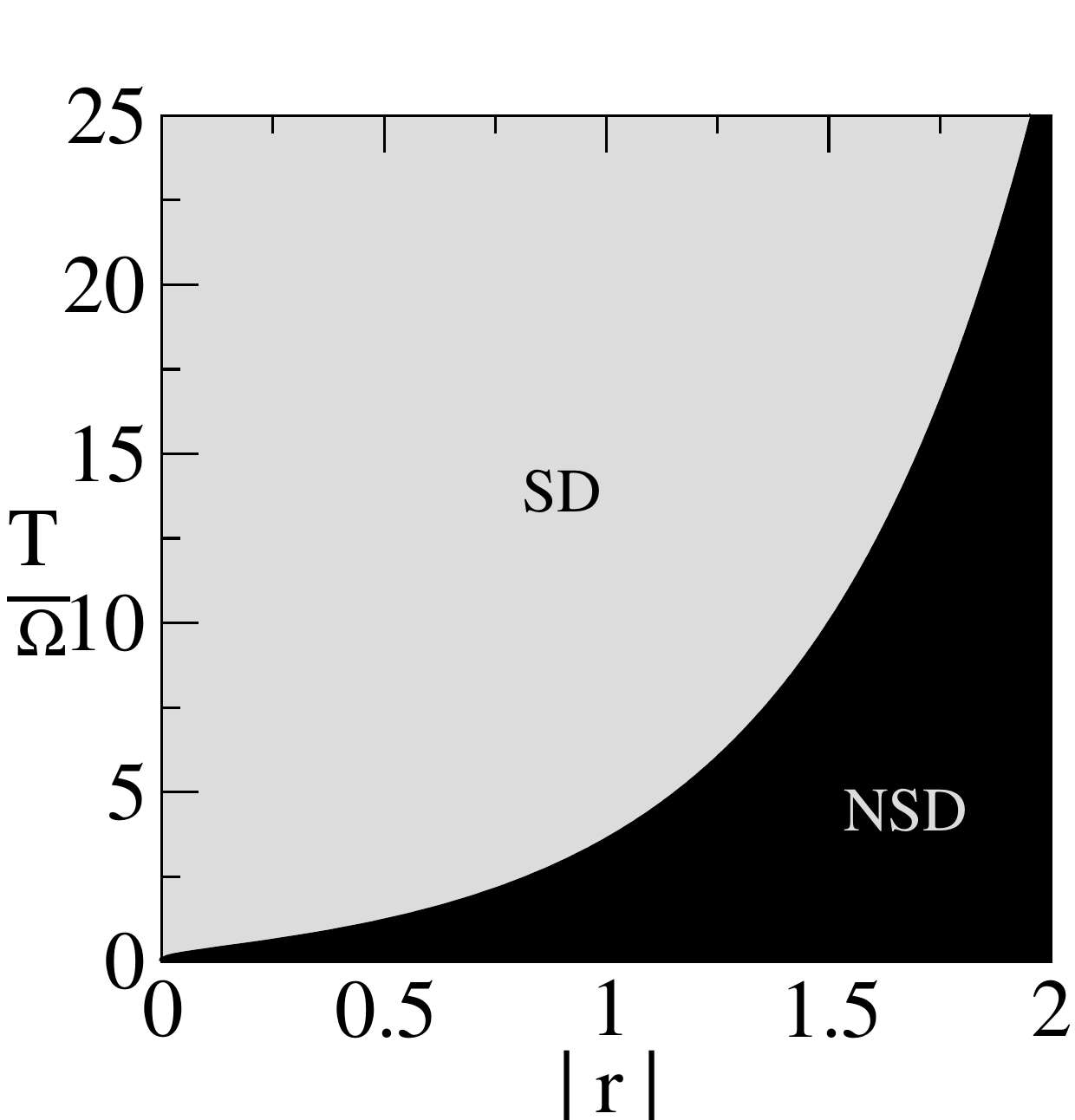}
\caption{Phase diagram for ohmic environment with symmetric coupling
 ($\Omega=1$, $M=1$, $C_{12}=0$, $\delta x_-\delta p_-=1/2$).
The phase diagram is qualitative different from the corresponding to the position coupling.
The SD and NSD phases are present. $r_{crit}=0$ for every temperature and the asymptotic entanglement
is allways constant. Here $E_{\mathcal N}$  is the distance from $|r|$ to the line that limits the
two phases.} 
\label{fig:phasesRWA}
\end{figure}

This implies that for symmetric coupling the phase diagram is simpler, as shown in Fig. \ref{fig:phasesRWA}. In this case there are only two phases (SDR does not exist). The NSD phase is characterized by the condition $|r|>\ln(4 \Delta x_+ \Delta p_+ \delta x_- \delta p_-)/2$. In this case, the entanglement achieved in the asymptotic regime is the difference between $|r|$ and the curve that limits the two phases. Contrary to what happens for position coupling, initial pure states (with $r=0$) belong to the SD phase at zero temperature. For a given value of the temperature $T$, the asymptotic state of has some degree of mixing. The states that have enough squeezing $r$ to support some entanglement for such degree of mixing are denoted as GLEMS (Gaussian least-entangled mixed states) \cite{Adesso04}. As a final comment we should point out that for an initial two-mode squeezed state, the curve defining the boundary of the NSD region (given by eq. (\ref{eq:EntRWA})) coincides with the one obtained in \cite{Prauzner04}, where entanglement 
was studied under a Markovian rotating-wave approximation.

\subsection{Evolution in the different phases: Analytic and numerical results.}
\label{sec:temp}

Here we will analyze the above results contrasting the analytic predictions with the results of an exact numerical solution of the problem. Numerical solution is indeed exact since in the case of a discrete environment (formed by $N$ oscillators). It is obtained by using the linearity of the problem to exactly evolve the complete covariance matrix and to obtain the full quantum state. Once this is done one can directly compute the logarithmic negativity (see \cite{Blume03,Blume07} for another application of the same method). Whenever possible (position coupling with ohmic spectral density) we compared this evolution with analytic expressions for the exact reduced evolution operator, finding complete agreement between both methods. 

\subsubsection{Position coupling}

\textit{a) Ohmic spectral density:}
Now, we will focus on the ohmic environment eq. (\ref{eq:mastereq})  $(n=1)$, where the high frequency cutoff $\Lambda$ defines a characteristic timescale $\Lambda^{-1}$ over which the coefficients $\gamma(t)$ and
$\delta\omega^2(t)$ vary. For times $t\gg\Lambda^{-1}$ these two coefficients settle into asymptotic values: $\gamma(t)\rightarrow\gamma=2\gamma_0$  
and $\delta\omega^2(t)\rightarrow -4\Lambda\gamma/\pi$. It is worth mentioning a technical point related with the renormalization that seems to have caused some confusion in the literature. The interaction with a common environment induces a coupling between the oscillators. Thus, even if we consider a vanishing "bare" coupling (i.e., $c_{12}=0$) the asymptotic value of the coupling will be non-zero and given by $C_{12}=\delta\omega^2/2$. It is natural to define renormalized parameters of the oscillators as the ones characterizing the long time limit. Thus, for the renormalized coupling to be $C_{12}=0$ we must consider a bare coupling $c_{12}=-\delta\omega^2/2$ in the original Hamiltonian. This simply says that the coupling constant between the oscillators must be renormalized in the same way as their natural frequency (with the same
counterterm). If one does not do this (and assume, for example, that the bare coupling vanishes) one would observe high frequency oscillations at long times (with a frequency which is set by the cutoff $\Lambda$). On the contrary, by adding the appropriate counterterms to the bare Hamiltonian one obtains a $\Lambda$--independent long time limit. In such case, we have $\Omega_1=\Omega_2=\Omega=\omega_-$.

\begin{figure}[htb!]
\vspace{0.2cm}
\includegraphics[width=8.5cm]{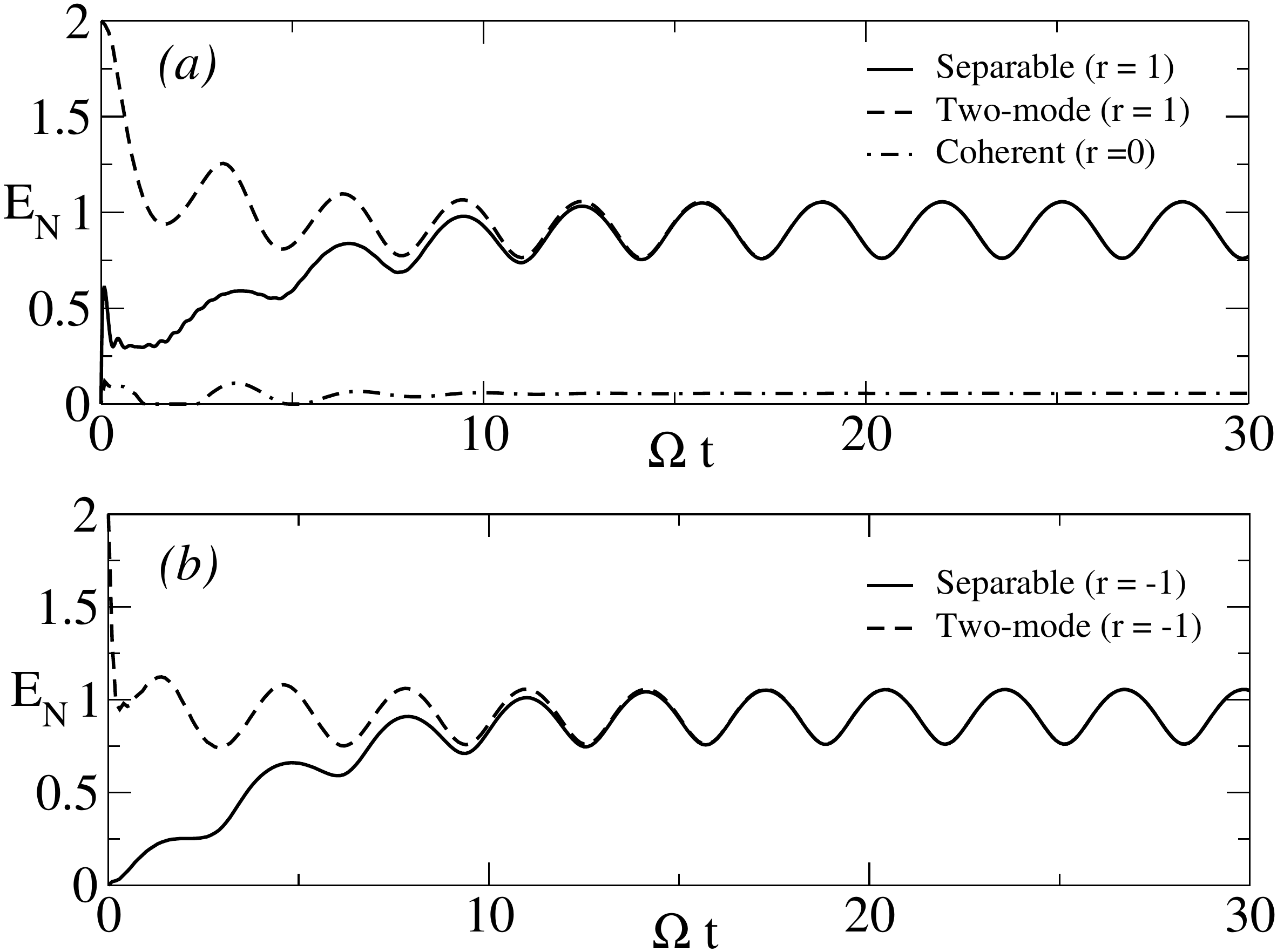}
\caption{Logarithmic negativity for resonant oscillators 
in the same environment. $(a)$ For $T=0$ the NSD phase appears both 
for large and small squeezing. Initially separable states, squeezed or coherent can get entangled.
The asymptotic behavior only depends on $r$. The amplitude of oscillations vanishes when $r\rightarrow 0$.
$(b)$ Initial states with negative squeezing, have the same asymptotic behavior with a dephasing
of $\pi/2$.}
\label{fig:entTime}
\end{figure}

Predictions discussed in the previous Sections can be verified by an exact numerical solution to the problem. For our numerics we considered parameters  $\gamma_0=0.1$, $\Omega=1$, $\Lambda=20$, $m=1$, $C_{12}=0$ (extension to the case where the natural oscillators interact can be easily done). We considered separable squeezed states for which $m\Omega\delta x_{1,2}/\delta p_{1,2}=\exp(2r)$ as well as two-mode squeezed states for which 
$m\Omega\delta x_+ /\delta p_+ =\delta p_- /(m\Omega\delta x_-) =\exp(2r)$ (in both cases $\delta x_- \delta p_-=1/2$). In Fig. \ref{fig:entTime}, we show
the entanglement dynamics in an environment at zero temperature. 
We clearly see that the final entanglement achieved by different initial states only depends  upon the squeezing factor $r$. Initial entangled states reduce their degree of entanglement while initial separable states do get entangled through the interaction with a common environment. 
Evolution of separable states with positive and negative squeezing is compared in Figs. \ref{fig:entTime} $(a)$ and $(b)$. In the first case entanglement grows much faster. This is due to the fact that the initial state has a wider spread in the position observable, which is the one appearing in the interaction Hamiltonian. 
In the asymptotic regime, as it is predicted, the dynamics is the same. They oscillate with the same frequency around the same mean value and with the same amplitude, but as it was expected, they have a phase shift of $\pi/2$. 

\begin{figure}[htb!]
\vspace{0.2cm}
\includegraphics[width=8.5cm]{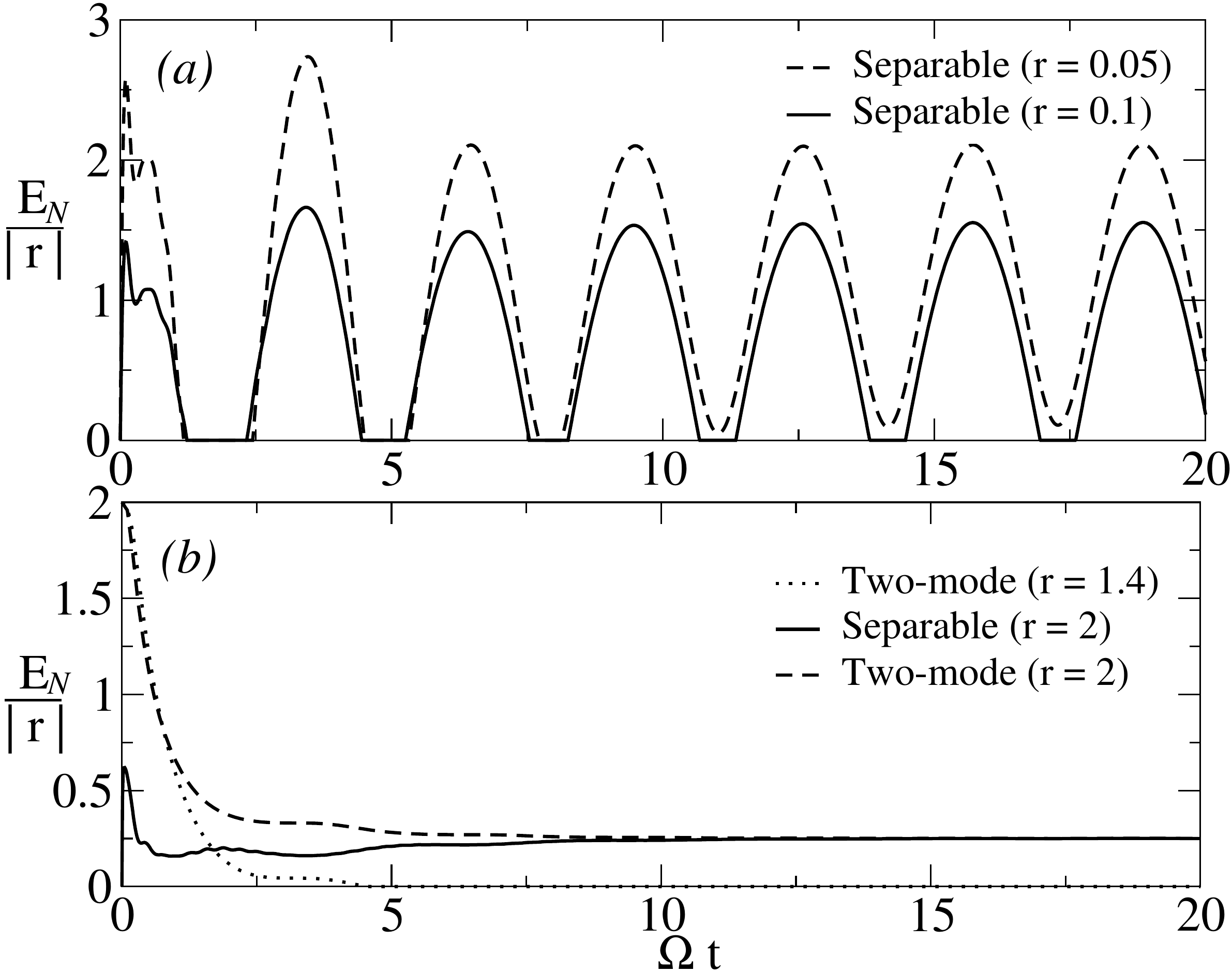}
\caption{$(a)$ The SDR phase appears for intermediate values of squeezing at zero temperature.
For $|r|<|r_{crit}|$ (dashed line) $\tilde E_{\mathcal N}=E_c$ and the amplitude of oscillations is equal to $|r|$.
$(b)$ $T/\Omega=10$, the SD phase appears for small $|r|$ and NSD phase for large squeezings, oscillations 
in the steady state are attenuated as the temperature increases.}
\label{fig:entTime2}
\end{figure}

The existence of events of sudden death and revival can also be seen from the 
numerical solution and are shown in Fig. \ref{fig:entTime2} (our numerical results show full agreement with the analysis presented above concerning the nature of the SDR phase). In the same Figure, we also show the evolution belonging to the NSD phase. They correspond to  a squeezing such that $|r|<|r_{crit}|$. In such case the amplitude of oscillations in the asymptotic regime is equal to $|r|$ and the mean value is $E_c$. An example of the SD phase appears 
in Fig. \ref{fig:entTime2} along with another example of the NSD phase for a  non-zero temperature. It is also noticeable that the amplitude of the 
oscillations almost vanishes in the high temperature limit.

We can obtain an analytical expression for the parameters needed to analyze the entanglement dynamics. Thus, using the exact expressions obtained in \cite{Fleming07} we find that at zero temperature

\beqa
&&r_1\equiv E_c(T=0)={1\over 2}\ln\Big[{\pi \over 2}{\sqrt{1-\gamma^2/\Omega^2}\over \arccos(\gamma/\Omega)}\Big], \nonumber \\
&&r_2={1\over 2}\ln\Big[{2-4 \gamma^2/\Omega^2 \over\sqrt{1-\gamma^2/\Omega^2}}\arccos(\gamma/\Omega)+{4\over \pi} {\gamma\over\Omega} \ln\Big[{\Lambda\over\Omega}\Big]\Big], \nonumber \\
&&r_{crit}={1\over 4}\ln\Big[1-2 {\gamma^2\over\Omega^2} +{2\gamma/\Omega \sqrt{1-\gamma^2/\Omega^2}}\frac{\ln[\Lambda/\Omega]}{\arccos(\gamma/\Omega)}\Big], \nonumber \\
&&S_{crit}=\frac{1}{4}\ln\Big[{4\over\pi^2} {1-2\gamma^2/\Omega^2 \over
{1-\gamma^2/\Omega^2}}\arccos^2(\gamma/\Omega)\nonumber \\
&&\ \ \ \   +{8\over\pi^2} {\gamma/\Omega \over \sqrt{1-\gamma^2/\Omega^2}}\ln\Big[{\Lambda\over\Omega}\Big]\arccos({\gamma/ \Omega})\Big]. \nonumber
\eeqa

These formulae have a simpler form in the weak coupling limit where 
\beqa
&&r_1\approx{1\over2}\ln\Big[1+{2\gamma\over\pi\Omega}\Big],	\nonumber\\
&&r_2\approx {1\over2}\ln\Big[{1}+\Big(\ln\Big[{\Lambda\over\Omega}\Big]-{1\over2}\Big){4\gamma\over\pi\Omega}\Big],\nonumber\\
&&r_{crit}\approx \frac{1}{4}\ln\Big[1+{4\over\pi}\ln\Big[{\Lambda\over\Omega}\Big]{\gamma\over\Omega}\Big],\nonumber\\
&&S_{crit}\approx {1\over4}\ln\Big[1+\Big(\ln\Big[{\Lambda\over\Omega}\Big]-1\Big){4\gamma\over\pi\Omega}\Big].\nonumber
\eeqa
In this case, the asymptotic coefficients of the master equation up to second order in $\gamma$ are given by: 
\beqa
D&\approx&m\gamma\Omega+{2m\gamma^2\over\pi}\Big(2\ln\Big[\frac{\Lambda}{\Omega}\Big]-1\Big),\nonumber\\
f&\approx&\frac{2 \gamma}{\pi}\ln\Big[\frac{\Lambda}{\Omega}\Big].\label{eq:fohmZT}
\eeqa 
A technical comment is in order here: To estimate the asymptotic behavior using an expansion in powers of the coupling constant, we need the coefficient $D$ to one order higher than $f$. This fact was already noticed in \cite{Fleming07} and is evident from the fact that critical squeezing is given by
\beq
r_{crit}={1\over 4} \ln\Big[1-{2m \gamma f\over D}\Big].
\label{eq:rcritFD} 
\eeq

Estimates for the critical temperature $T_0$ (the temperature for which the position dispersion becomes identical to the vacuum one) can be obtained as follows: position dispersion at low temperatures is:
\beqa
\Delta^2x(T)=\frac{T}{\Omega^2m}+\frac{1}{\pi m \sqrt{\Omega^2-\gamma^2}}
\rm{Im}\Big[\rm{H}\Big(\frac{\gamma+i\sqrt{\Omega^2-\gamma^2}}{2 \pi T}\Big)\Big],\nonumber
\eeqa
where the function $H(z)$ is the Harmonic Number. Expanding this for low temperatures ($T/\Omega\ll1$),
\beqa
&&m\Omega\Delta^2x(T)
 \approx{\arccos(\gamma/\Omega)
 \over\pi\sqrt{1-(\gamma/\Omega)^2}}+
{2\pi\over3} \gamma/\Omega\Big({T\over\Omega}\Big)^2\nonumber\\
&&+{8\pi^3\over15}(1-2(\gamma/\Omega)^2)\Big({T\over\Omega}\Big)^4,
\eeqa
we can obtain an approximate expression for $T_0$ that accurately reproduces our results for $\arctan\big({\sqrt{1-\gamma/\Omega)^2}/(\gamma/\Omega)}\big)\ll\pi/2$.

On the other hand, for high temperatures we can use the appropriate approximations to obtain:
\beqa
&&r_{crit}\approx \frac{1}{4}\ln\Big[1+\frac{2 \gamma}{\pi \Omega}\ln\Big[\frac{\Lambda+\Omega}{\Lambda-\Omega}\Big]\Big],\nonumber\\
&&S_{crit}\approx {1\over2}\ln\Big[2 {T\over\Omega}\Big]+{1\over4}\ln\Big[1+\frac{2 \gamma}{\pi \Omega}\ln\Big[\frac{\Lambda+\Omega}{\Lambda-\Omega}\Big]\Big].\nonumber
\eeqa
In this regime $r_{crit}$ approaches a temperature-independent value that decreases with the high frequency cutoff and increases with the
coupling constant $\gamma$. As a consequence, the asymptotic entanglement is approximately constant. The behavior of $S_{crit}$ is simpler: as expected it behaves as the entropy, growing as $\ln(T)$ 
for high temperatures. The narrow passage between the SD and the NSD phases closes as $1/\Lambda$ 
and moves to larger and larger values of squeezings. For completeness we include the diffusion coefficients
in the high temperature regime up to first order in $\gamma$. They are 
\beqa
D&\approx &2m\gamma T,\nonumber\\
f&\approx& -\frac{2 \gamma}{\pi\Omega}
\ln\Big[\frac{\Lambda+\Omega}{\Lambda-\Omega}\Big]T.\label{eq:fohmHT}
\eeqa

\textit{b) Sub-ohmic spectral density:}
Here we will analyze the behavior of entanglement in an environment with a sub-ohmic spectral density as (\ref{eq:spectral}) with  $n=1/2$. In this case, the oscillators of the infrared and the resonant bands are coupled more 
strongly to the system (since $\Omega \ll\Lambda$) and the environment induces more dissipation. As a consequence, the equilibrium state of the oscillator $x_+$ is noticeably more squeezed along position than the one corresponding to the ohmic case \cite{PazRoncaglia08}.
Therefore, considering eq. (\ref{eq:deltaE}), we expect a larger value for $|r_{crit}|$ at zero temperature which, 
in turn, would imply that the oscillations of the entanglement in the steady state will have 
larger amplitude.  In addition, the entanglement for initial coherent states will be larger as well as the critical temperature $T_0$. Below we will show only the numerical results, since there are no available analytic expression for the coefficients of the master equation. We used the same parameters as in the previous subsection, 
noticing that in this case $\delta \omega^2(t)\rightarrow-8(2 \gamma_0)/\pi\Lambda$.  

\begin{figure}[htb!]
\vspace{0.2cm}
\includegraphics[width=8.5cm]{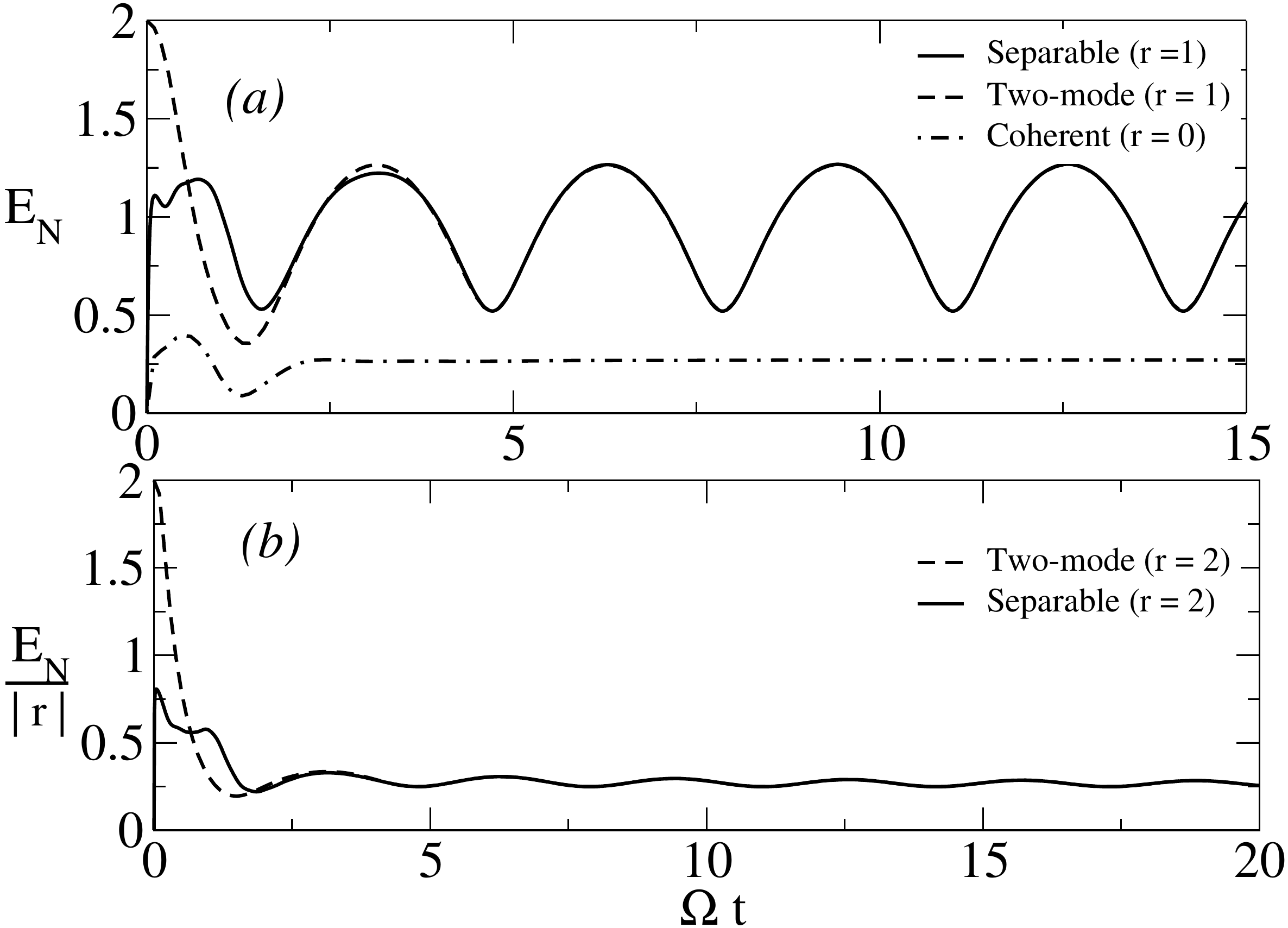}
\caption{Logarithmic negativity for resonant oscillators 
in a sub-ohmic environment. $(a)$ For $T=0$ the NSD phase appears both 
for large and small squeezing. The amplitude of oscillations are higher than
in the ohmic case. And as a consequence the entanglement achieved for initial coherent
states is also higher. $(b)$ For $T/\Omega=10$ appreciable oscillations are present in the asymptotic
regime.}
\label{fig:entTime3}
\end{figure}

In Fig. \ref{fig:entTime3}, we show the dynamics of entanglement for two resonant oscillators 
immersed in a bath at zero temperature. There we can appreciate oscillations
of entanglement in the steady state with larger amplitude than in the ohmic case but with the same frequency.
For initial coherent states, the entanglement achieved is greater than in the ohmic case. 
As mentioned above, this is a consequence of the coupling 
between the system and the resonant bands of the environment that produce a substantial squeezing in the steady state. 
We can also notice that the system approaches equilibrium earlier than in the ohmic case due to the fact that dissipation is stronger than in the ohmic case. 

\begin{figure}[htb!]
\vspace{0.8cm}
\includegraphics[width=8.7cm]{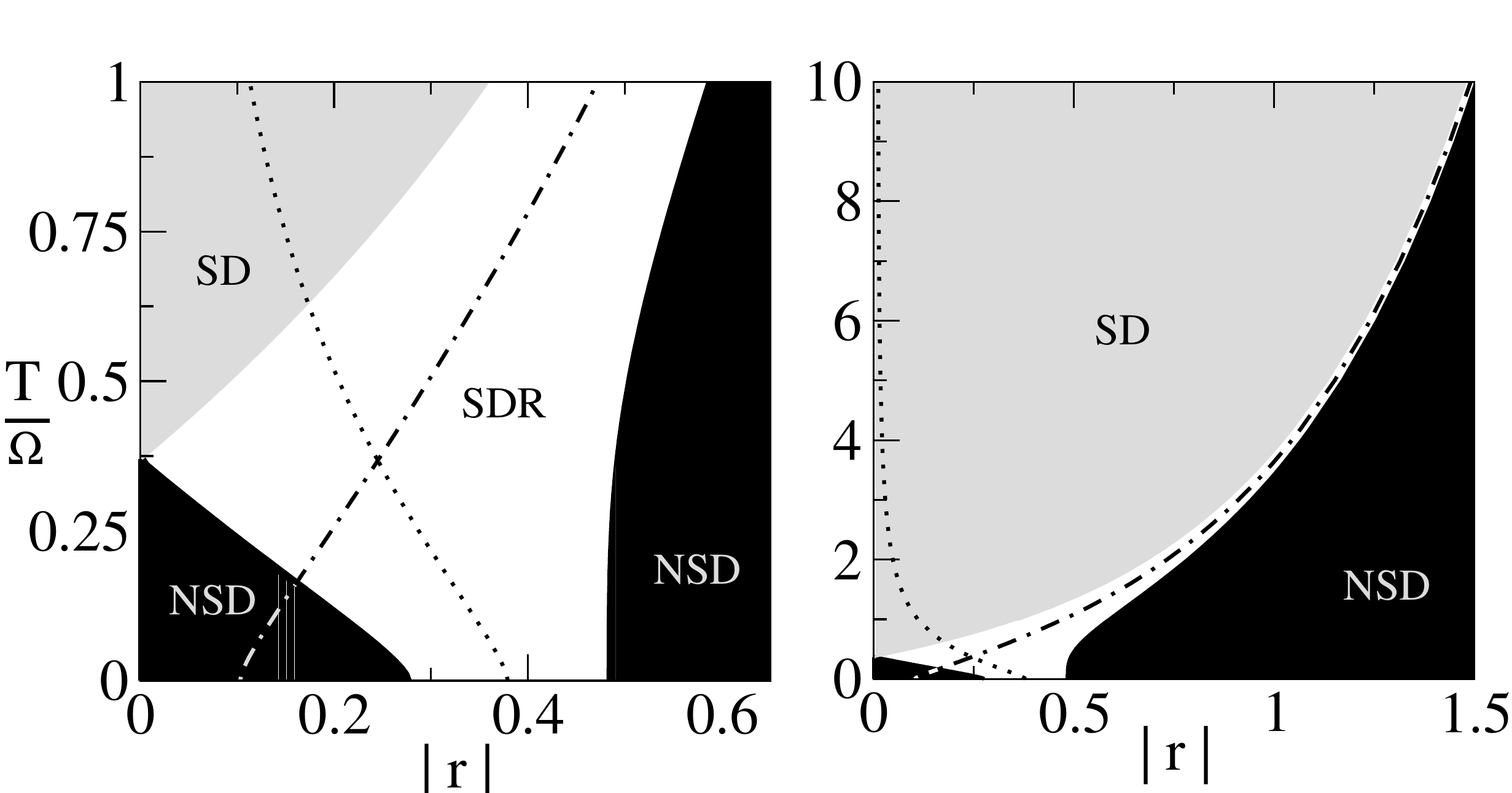}
\caption{Phase diagram for sub-ohmic environment.
The SD, NSD and SDR phases that describe the three different qualitative long time behaviors
for the entanglement are present. The low temperature NSD island is bigger than 
the one corresponding to the ohmic environment.} 
\label{fig:Subphases}
\end{figure}

The asymptotic features of the entanglement observed in the figures for specific cases, can be 
summarized in the phase diagram shown in Fig. \ref{fig:Subphases}. 
The shape of this diagram is essentially the same as the preceding case. In the low temperature regime we find again an NSD island with an area that is larger than the one corresponding to the ohmic case. 
Also the value of $|r_{crit}|$ at zero temperature is larger than for the ohmic environment, and it decreases with the temperature. 
As a consequence, we also observe a SDR region at high temperature whose width is given by $|r_{crit}|$. In this case, the oscillations that appear in the ohmic spectral density are enhanced.

Even though there are no analytic expressions for this environment, we can obtain approximate formulae in the weak coupling and high cutoff regime. We can use them to get some qualitative understanding of the expected behavior but they are useles to draw quantitatively conclusions (since we can only obtain all coefficients of the master equation up to first order in $\gamma_0$ but, as indicated above, to estimate quantities such as $r_{crit}$ and $S_{crit}$ we would need to have the asymptotic 
form of the coefficient $D$ to one order higher).	
Thus, to lowest order we find  $\gamma(t)\rightarrow\gamma_{sub}=2\gamma_0\sqrt{\Lambda/\Omega}$, which grows with the cutoff as expected. At zero temperature, the diffusion coefficients are
\beqa
D&\approx& m \gamma_{sub}\Omega, \\
f&\approx&\gamma_{sub}\Big(1-{2\over\pi}\sqrt{\Lambda\over\Omega}	\ln\Big[{\Lambda+\Omega\over\Lambda-\Omega}\Big]\Big).\label{eq:fsubZT}
\eeqa
Again, the two coefficients grow with the coupling and the cutoff frequency.
The anomalous diffusion coefficient $f$ is larger than the one corresponding to the ohmic case. This produces a stronger localization of the asymptotic state in the position observable. We can also obtain the high temperature expressions in the weak coupling limit:
\beqa
D&\approx& 2m\gamma_{sub}T,\\
f&\approx&-2\gamma_{sub}\frac{T}{\Omega}.\label{eq:fsubHT}
\eeqa 
In this case all the coefficients are proportional to the temperature. The fact that the asymptotic state is squeezed at high temperatures is a consequence of the fact that the coefficient $f$ approaches significantly higher values than the ones corresponding to the ohmic case.

\begin{figure}[htb!]
\vspace{0.5cm}
\includegraphics[width=8.5cm]{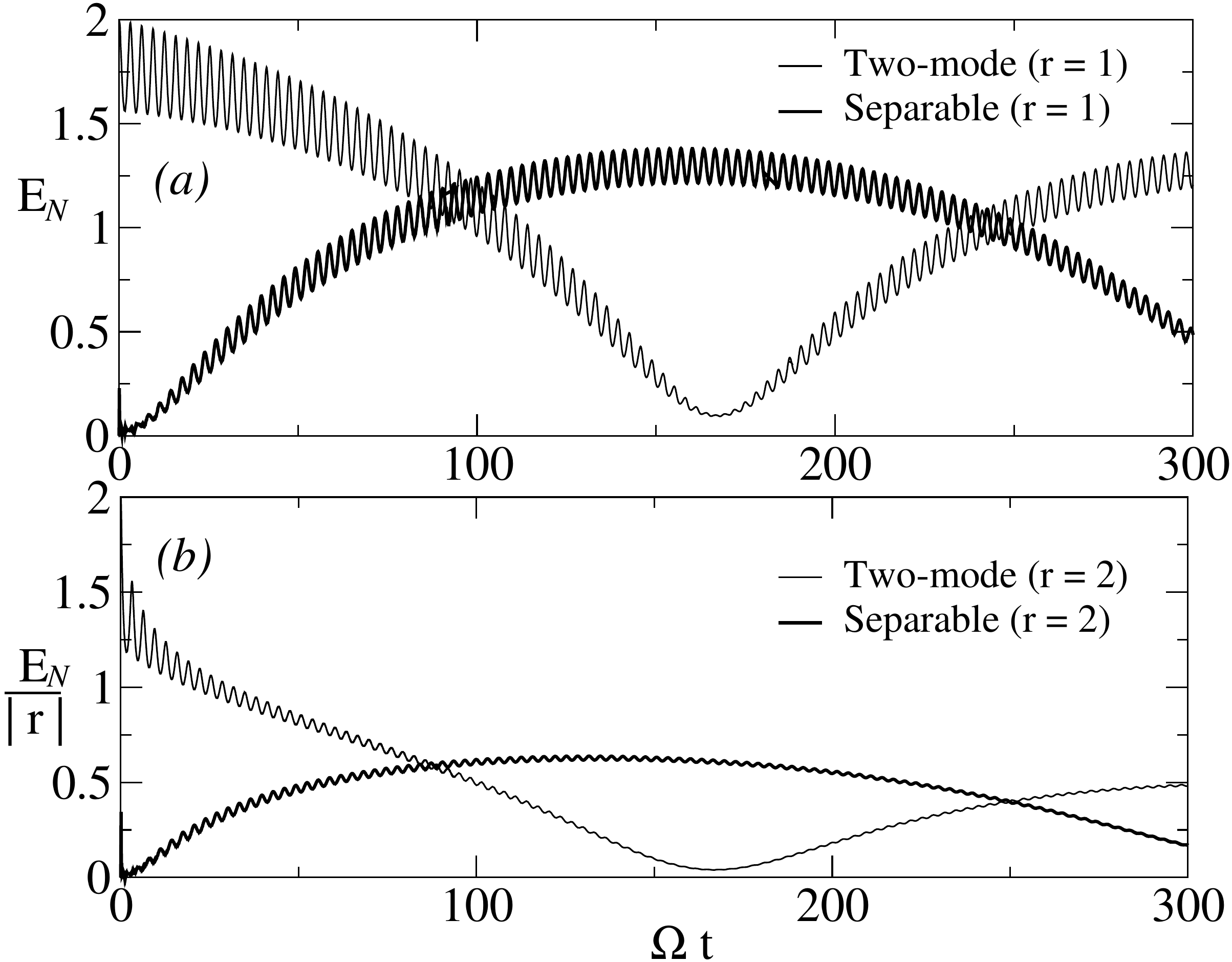}
\caption{Logarithmic negativity for resonant oscillators 
in a super-ohmic environment with $\gamma_0=0.15$. (a) For $T=0$ we do not observe that the entanglement
achieves equilibrium. Oscillations are present for long times. 
(b) At $T/\Omega=10$ the entanglement oscillates with smaller amplitude.}
\label{fig:entTimeSuper}
\end{figure}

\textit{c) Super-ohmic spectral density:}
A super-ohmic environment has a spectral density characterized by a higher population of high frequency bands. A typical example is given by equation (\ref{eq:spectral}) with $n=3$. Super-ohmic environments are weakly dissipative. In fact, in this case the dissipation coefficient approaches an asymptotic value given by $\gamma(t)\rightarrow\gamma_{sup}=2\gamma_0(\Omega/\Lambda)^2$. The frequency shift is $\delta \omega^2(t)\rightarrow-4(2 \gamma_0)\Lambda/3\pi$.
Thus, dissipation strictly vanishes in the infinite cutoff limit. In such case the oscillator $x_+$ does not reach equilibrium, a fact that was also noticed in \cite{PazHabZur93} and is related with the phenomenon of recoherence that could be induced by this type of environment (i.e., decoherence is reversible in this case). In this limit we cannot apply the analysis presented in the previous Sections, which requires the oscillator $x_+$ to approach equilibrium. Thus, for a super--ohmic environment we expect to observe an oscillatory behavior for the entanglement up to very long times. This is precisely what is observed in Fig. \ref{fig:entTimeSuper}, where we show the results of the numerical solution for two different initial states. Oscillations of entanglement persist for low and high temperatures. The amplitude of the oscillations decreases very slowly with time simply due to the fact that we consider a finite value for the high frequency cutoff. For this reason the value of the dissipation coefficient is not strictly zero but very small. This implies that the system would reach an equilibrium in the extremely long time limit (i.e. for times of the order of $1/\gamma_{super}$, an estimate which is consistent with the numerically observed behavior). It is worth noticing that the result we present here for the super--ohmic environment is not compatible with the ones reported in \cite{An07} (the super-ohmic results of that paper seem to be simply in error, a more detailed comparison with such results will be presented below). 

For zero temperature we can also obtain the asymptotic behavior of the diffusive coefficients in the weak coupling limit. These coefficients behave as  $f\approx{2\gamma_0/\pi}+{\gamma_{sup}}\ln[{(\Lambda^2-\Omega^2)/\Omega^2}]/\pi\approx{2\gamma_0/\pi}$ and $D\approx m\Omega\gamma_{sup}$. Thus, in this case the anomalous diffusion $f$ is proportional to the coupling constant and becomes independent of the cutoff. It takes the smallest value, comparing the three spectral densities that
we considered, which is a signature of the weak coupling between the system and the resonant band of the environment. On the other hand, $D$ vanishes in the infinite cutoff limit (as mentioned above, $\gamma_{sup}$ vanishes as well). In the high temperature regime we have $f\approx2\gamma_0 T/\pi\Lambda$ and $D\approx 2mT\gamma_{sup}$. Here the small value of $f$ produces a squeezing of the asymptotic state which is smaller than the one achieved for ohmic and sub-ohmic environments.

\subsubsection{Coupling symmetric in position and momentum}

Here we will consider the case where the coupling to the environment is symmetric in position and momentum. This model at zero temperature was studied previously in \cite{An07} with a two-mode squeezed state as the initial condition. Here we extend these results by considering arbitrary initial Gaussian states and arbitrary temperatures (we also take the opportunity to correct some erroneous results reported in \cite{An07}). The main conclusion concerning entanglement dynamics was announced before: only two phases (NSD and SD) exist. This conclusion follows from the fact that the master equation is symmetric under canonical interchange between position and momentum. It is independent of the precise form of the asymptotic values of the coefficients appearing in the equation as long as equilibrium exist (which is not the case for the super-ohmic environment).  

\begin{figure}[ht]
\vspace{0.5cm}
\includegraphics[width=8.5cm]{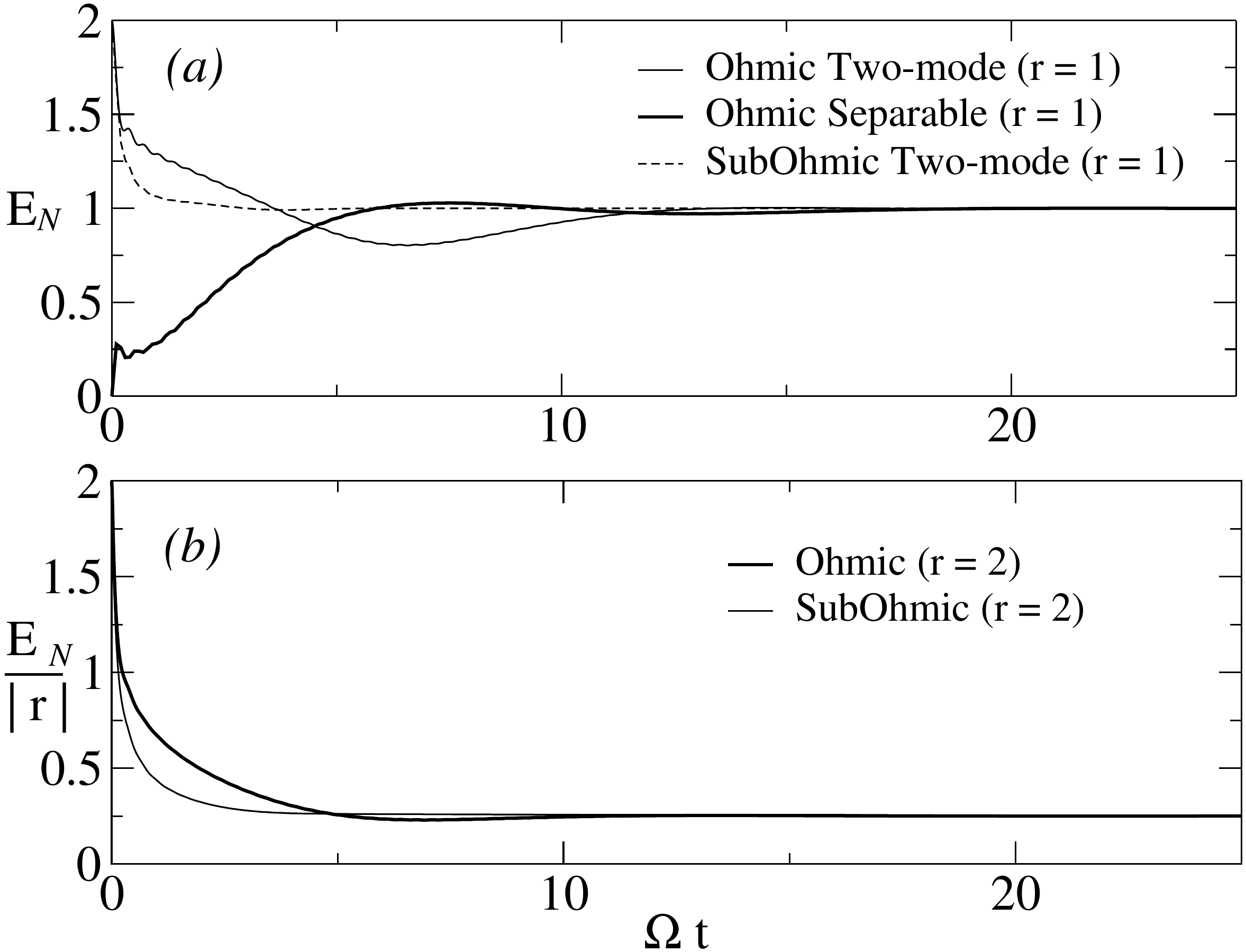}
\caption{Entanglement dynamics for resonant oscillators in an environment with symmetric coupling. 
$(a)$ Environment at zero temperature, the asymptotic entanglement depends on the squeezing $r$ and is constant. Ohmic and sub-ohmic environments arrive at the same equilibrium entangled state.
$(b)$ Environment at $T/\Omega=10$, the final entanglement depend upon the initial squeezing for
both spectral densities.}
\label{fig:timeRWA}
\end{figure}

We confirm this by a detail study of the numerical solution using the same parameters of the previous sections. Here we also considered $C_{12}=\tilde C_{12}=0$ (then, $\Omega=\omega_-$ and $M=m_-$). In Fig. \ref{fig:timeRWA} we show the dynamics of entanglement for ohmic and sub-ohmic spectral densities. Our result show that at zero temperature entanglement is reduced to exactly half of its initial value \cite{An07}. This is a prediction of eq. (\ref{eq:ent}) which is valid both for ohmic and sub-ohmic spectral densities, since in all those cases the asymptotic state of the oscillator $x_+$ is the ground state. In fact, the form of the master equation at zero temperature ensures the stability of the ground state.  As the asymptotic state is pure then this process can be thought of as a way to create pure gaussian entangled states from initial separable ones. Another obvious consequence of the symmetric coupling is that the behavior of initial states with negative or positive squeezing is identical. In Fig. \ref{fig:timeRWA} we show an example of the behavior of entanglement at temperature different from zero. As we discussed above, the steady state has non--zero entanglement. These results are summarized in the simple phase diagram of Fig. \ref{fig:phasesRWA}, which is essentially the same both for ohmic and sub-ohmic environments.

\begin{figure}[ht]
\vspace{0.5cm}
\includegraphics[width=8.5cm]{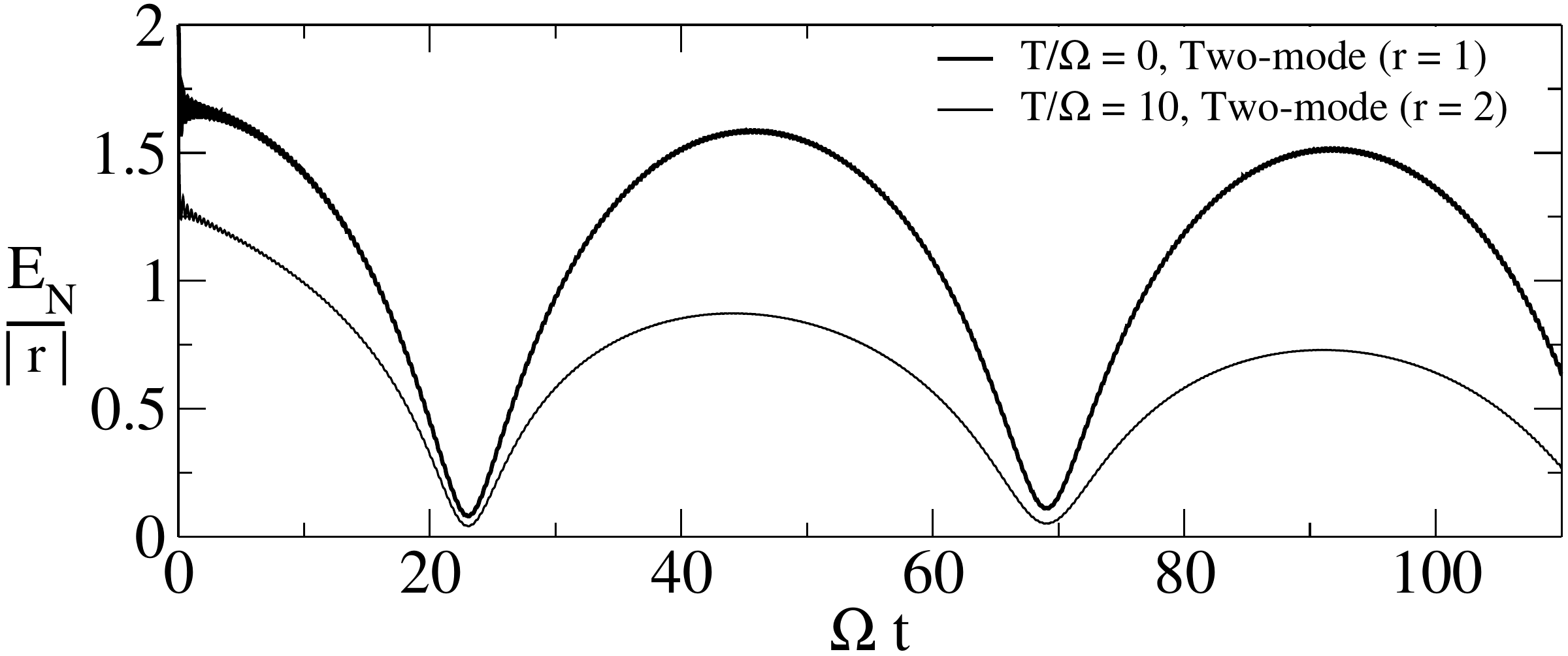}
\caption{Oscillators with symmetric coupling immersed in a super-ohmic environment.
$T=0$, the entanglement oscillates approaching approximately its initial value.
The amplitude of oscillations decreases slowly since we are considering a finite cutoff.
We also observe oscillations of entanglement for higher temperatures.}
\label{fig:timeSuperRWA}
\end{figure}

For a super-ohmic environment the  dissipative coefficient $\tilde \gamma$ scales as $1/\Lambda^2$. In Fig. \ref{fig:timeSuperRWA} we show that entanglement oscillates for very long times, a simple consequence of the vanishingly small value of the dissipative coefficient. This contradicts the results obtained in \cite{An07} where it was shown that for a super-ohmic environment entanglement achieves equilibrium before the ohmic and sub-ohmic cases. From our previous analysis, based on the use of the master equation, we can simply conclude that the results of \cite{An07} do not seem to be reliable. On the contrary, our numerical results support the simple conclusion obtained analytically by using the master equation, which is local in time. Thus, entanglement oscillates slowly decaying with a rate that is roughly given by $\tilde\gamma$ (which goes to zero in the  infinite cutoff limit). 

It is simple to obtain analytic estimates for the asymptotic values of the time dependent coefficients using a perturbative approximation. In that case, the dissipation constant is $\tilde \gamma\rightarrow 4\gamma_0(\Omega/\Lambda)^{n-1}=2 J(\Omega)\pi/\Omega m$. This shows that sub-ohmic environment induces stronger dissipation than an ohmic one. Also, for  supra-ohmic environments ($n>1$) there is not equilibrium in the infinite cutoff (Markovian) limit.  In the same way, we can obtain the diffusion coefficient which is given by $\tilde D=2J(\Omega) \pi\coth({\Omega\over2 T})$. Using this expression the asymptotic dispersions for the $x_+$ oscillator are:
\beq
M^2\Omega^2\Delta x_+^2=\Delta p_+^2={M\Omega\over 2}\coth\Big({\Omega\over2 T}\Big),
\label{eq:dispSym}
\eeq
These approximate expressions enable us to recover the results reported in  \cite{Prauzner04}.

\section{Non-resonant oscillators}
\label{sec:offres}

The above properties are valid under a single important assumption: 
the two oscillators are resonant. If this is not the case the analysis becomes more 
complicated. The master equation is no longer valid since the $x_\pm$ modes 
are coupled. As $x_-$ is not isolated it also approaches equilibrium. 
To analyze this we can obtain a new perturbative master equation (assuming the interaction is through the position). 
It reads:
\begin{widetext}
\beqa
\dot\rho&=&-i[H_R,\rho] -i\gamma(t)[x_+,\{p_+,\rho\}]-D(t)[x_+,[x_+,\rho]]
-f(t)[x_+,[p_+,\rho]] \nonumber \\
&-&i{m\over2}\delta\Omega_{+-}^2(t)[x_+,\{x_-,\rho\}]-i\gamma_{+-}(t)[x_+,\{p_-,\rho\}]-D_{+-}(t)[x_+,[x_-,\rho]]
-f_{+-}(t)[x_+,[p_-,\rho]].\label{eq:MEoffresonant}
\eeqa
\end{widetext}
As seen in the above equation, the $x_\pm$ oscillators interact with a coupling constant $c_{+-}=(\omega_1^2-\omega_2^2)/2$ while $x_+$ is directly coupled to the environment. 
One of the terms coupling $x_\pm$ in (\ref{eq:MEoffresonant}) is a  renormalization of the coupling constant. There is also a diffusive and a dissipative term. All the coefficients labeled with $\pm$ indices are proportional to the detuning $\Delta=(\omega_1-\omega_2)$ (all of them vanish in the resonant limit). We can obtain asymptotic dispersions of the two oscillators, but the corresponding formulae are rather cumbersome.  The most important generic conclusions that we can draw from the above non-resonant master equation are the following. 1) As there is a final equilibrium state for both $\pm$ oscillators, the final entanglement becomes independent of the initial state. 2) The approach to equilibrium proceeds with two different timescales (one of the decay rates is proportional to the detuning). 3) For sufficiently high temperatures the generic fate of the asymptotic regime is SD and the final state contains no entanglement. 4) However, for very low temperatures the final state can be entangled. The origin of the final entanglement resides again in the squeezing of the equilibrium state. If the $\pm$ oscillators reach a final state with different squeezing (the squeezing of the $x_+$  oscillator is different from that of $x_-$ mode) then the final state may be entangled.  For very low temperatures this condition may be verified and entanglement may be present in the final state. 
\begin{figure}[htb!]
 \vspace{0.1cm}
\includegraphics[width=8cm]{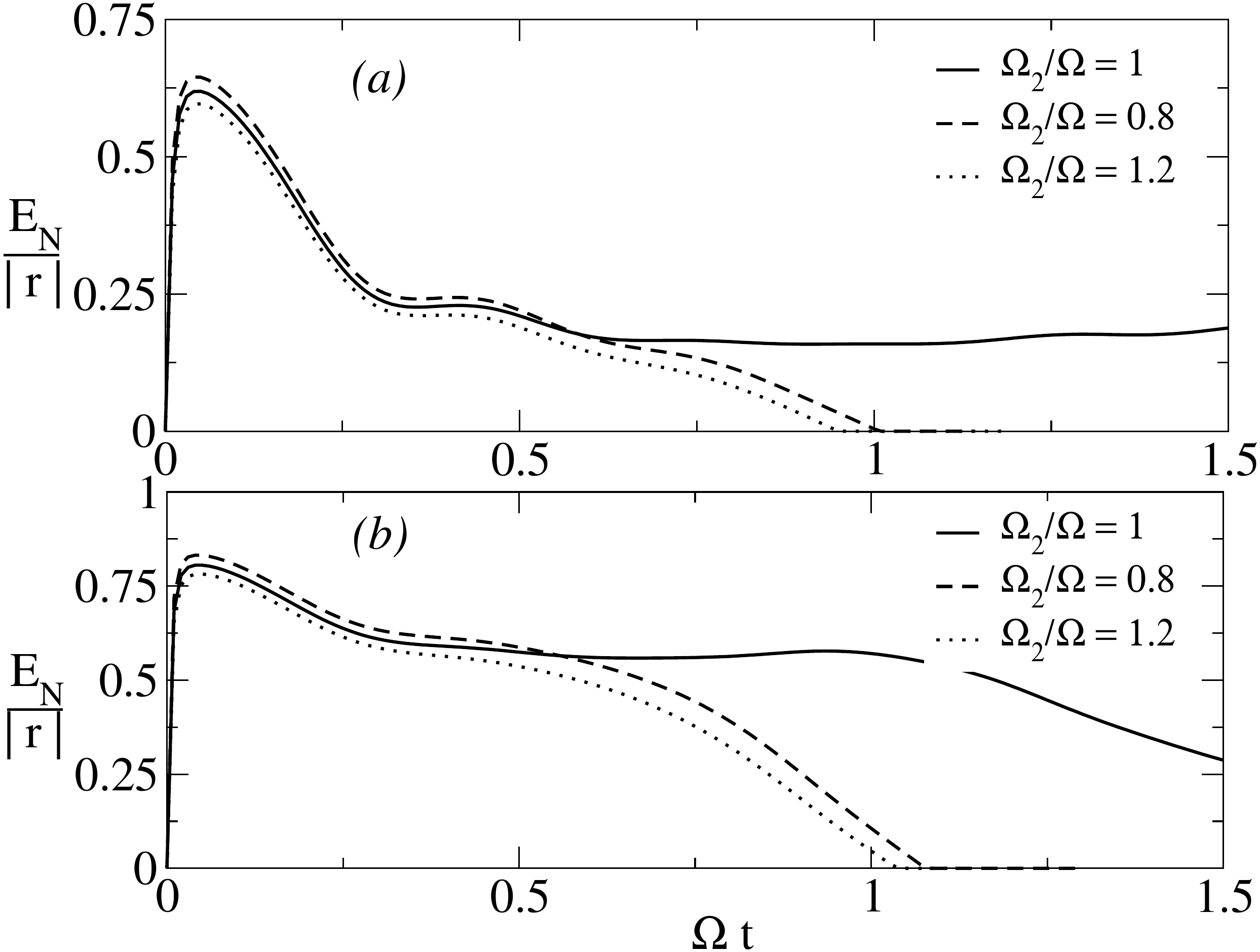}
\caption{Dynamics of entanglement for oscillators in an initial separable state $r=2$.
The dynamics of non-resonant oscillators along with the resonant case is shown for 
(a) ohmic  and (b) sub-ohmic environments at $T/\Omega=10$ with $\gamma_0=0.1$.}
\label{fig:FreqTime}
\end{figure}
In Fig. \ref{fig:FreqTime} we show how things change when we move away from the resonance condition if the environment is in a finite temperature state ($T/\Omega=10$). There we show the dynamics for ohmic and sub-ohmic spectral densities. For early times, initial separable states become entangled due to the action of the environment. However, entanglement decays much faster for non-resonant oscillators and the state becomes disentangled in a finite time (SD). 
We can also observe that the sub-ohmic environment can retain entanglement a bit longer than the ohmic one. This is due to the fact that the bare coupling (which is later cancelled by the coupling induced by the environment) produces more entanglement at short times in the sub-ohmic case. One can also notice little differences between the evolution corresponding to non-resonant oscillators with higher (or lower) frequencies. Indeed, this is due to the fact that the virtual interaction, $c_{+-}$, depends on the square of the frequencies and not upon the difference between them.
As both virtual oscillators approach an equilibrium state (which is characterized also by non-vanishing correlations between them) the final entanglement turns out to be independent of the initial state. Thus, generic fate of entanglement at sufficiently high temperature is sudden death. 

\begin{figure}[htb!]
 \vspace{0.2cm}
\includegraphics[width=8.7cm]{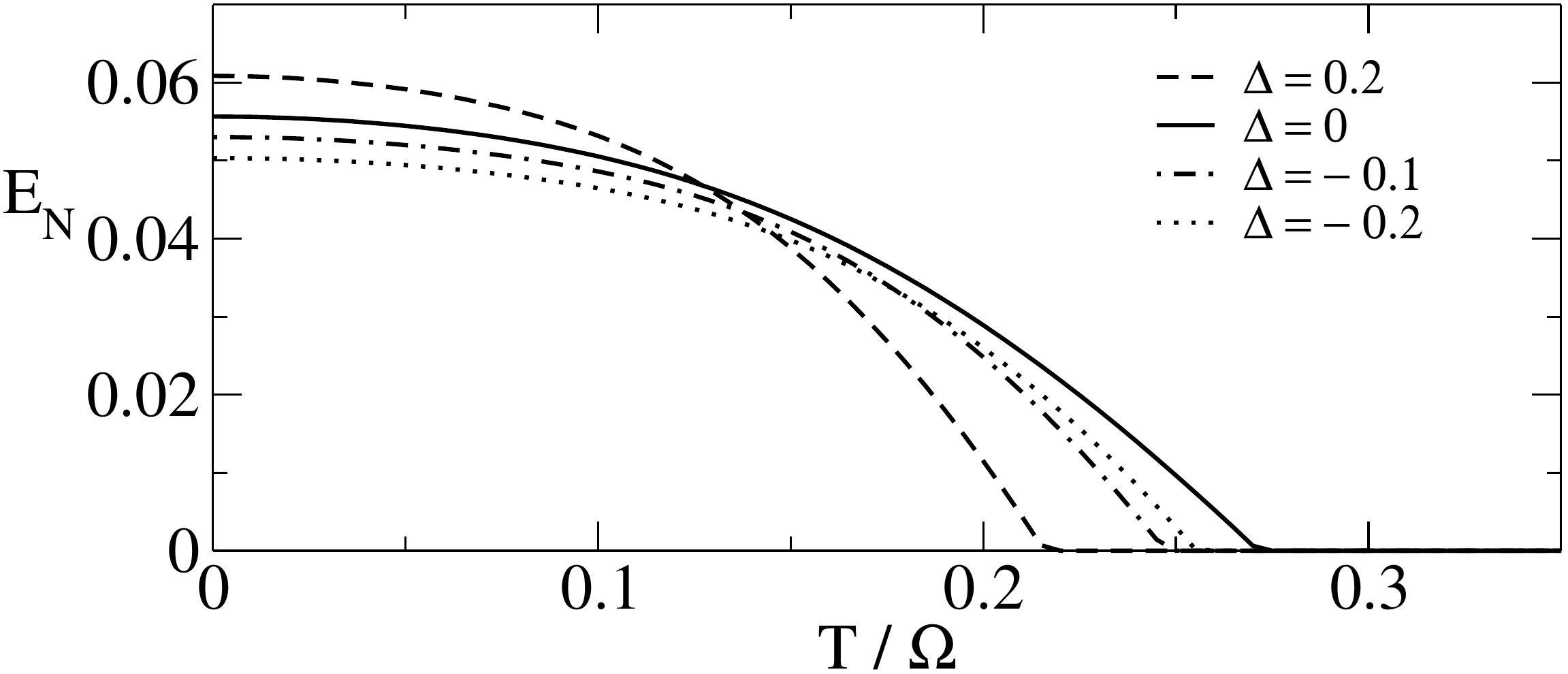}
\caption{Asymptotic entanglement between non-resonant oscillators as a function of the temperature for
different detunings $\Delta=\omega_1-\omega_2$ (ohmic spectral density). 
We can find a critical temperature for every detuning below which the state is entangled. 
They are compared with the entanglement for $r=0$ and $\Delta=0$.
}
\label{fig:EntTemp}
\end{figure}
It is interesting to notice that for any detuning it is possible to find temperatures below which the asymptotic state is entangled. The origin of this asymptotic entanglement, as mentioned above, lies on the final squeezing of the $x_\pm$ modes. The dependence of the final entanglement with temperature is analyzed in Fig. \ref{fig:EntTemp}. The curve is reminiscent of a phase transition with critical temperature depending upon the detuning. The existence of entanglement in the asymptotic state is not really a total surprise and is clearly related to recent findings of the existence of entanglement in the ground state of harmonic chains similar to the one we studied here \cite{Anders08}.

\section{Conclusions}
\label{sec:conc}

We presented a complete study of the evolution of the entanglement between two oscillators interacting with the same environment. We extended the analytical and numerical results previously presented in \cite{Paz08}. We considered two related models for the interaction between the system and the environment: one where the coupling is through position and another where the coupling is symmetric in position and momentum. In both cases we used an exact master equation as our main analytical tool. For position coupling we presented a phase diagram valid for ohmic and sub-ohmic spectral densities, and we showed that it contains three phases (SD, NSD and SDR). For both spectral densities the phase diagram is qualitatively the same. The main difference is that the sub-ohmic environment tends to enhance the amplitude of the entanglement oscillations (which is due to the fact that the asymptotic state induced by a sub-ohmic environment has larger squeezing than the one corresponding to the ohmic case). On the other hand, we showed that a qualitatively different phase diagram emerges when the coupling is symmetric. In that case, the SDR phase is 
absent and the asymptotic entanglement does not oscillate. Our results clearly show that initial separable states can get entangled and that initially entangled states can suffer from sudden death. 

For position coupling, we showed that there is a range of temperatures where SD never occurs. In fact, this is the case for $T\le T_0$ where $T_0$ is the temperature where the position dispersion of the $x_+$ oscillator becomes identical to the one corresponding to vacuum (below $T_0$ such dispersion is smaller due to squeezing).  On the other hand, for symmetric coupling the SD phase is present for every temperature. 

Our results can be extended in several ways. In fact, along the paper we have focused on the case where the renormalized oscillators do not interact, but our analysis can be applied to the cases where $C_{12}$ does not vanish. In this case, as $\Omega_+\neq\Omega_-$, the boundaries between different phases change slightly but the phase diagram remains qualitatively the same. For example, the formula for $r_{crit}$ given in(\ref{eq:rcrit}) tells us that when $\Omega_+\neq\Omega_-$  
$r_{crit}$ may be non-zero even if the state of the $x_+$ oscillator is not squeezed. In this case the resource for the asymptotic entanglement is, not surprisingly, supplied by the interaction. On the other hand, when the coupling is symmetric the asymptotic evolution does not change considerably by adding an interaction between the oscillators. Indeed, in this case, we always have $M\Omega=m_-\omega_-=m\omega$. Therefore, $r_{crit}$ vanishes and there are no entanglement oscillations in the long time regime. There is only one exception to this rule: If one introduces a non-symmetric coupling between the oscillators, i.e. $c_{12}\neq\tilde c_{12}$, then we get $M\Omega\neq m_-\omega_-$. 
Our results also change if the initial states of the system is mixed. However, the change in the phase diagram is simple to understand. In fact, the degree of purity of the initial state is characterized by the product $\delta x_- \delta p_-$, which only enters in the expressions of $S_{crit}$ and changes the mean value of the final entanglement (as seen in eqs. (\ref{eq:ent}) and (\ref{eq:Scrit})). It is simple to see that the entanglement achieved for pure sates is grater than the one obtained for mixed
initial states with the same degree of squeezing. The phase diagram for mixed states can be simply obtained from the one corresponding to pure states by shifting the curve $S_{crit}$ to the right. This has the effect of moving upwards the horizontal axis (see Fig. \ref{fig:phases}). As a consequence, the value of $T_0$ changes and the low temperature NSD island may disappear depending on the degree of impurity of the initial state.

The existence of asymptotic entanglement between resonant oscillators $x_1$ and $x_2$ can be understood in terms of the following quantum-optical analogy: We can think of these oscillators as two modes of the electromagnetic field. The evolution of such modes, interacting with the environment is equivalent to the following sequence of operations: i) a 50/50 beam splitter supperpose both modes (creating the $x_\pm$ oscillators out of the original ones), ii) while one of the output modes ($x_-$) evolves freely, the other is replaced by a new one with dispersions along its quadratures given by the equilibrium values (this operation entirely replaces the interaction between $x_+$ and the environment), iii) another 50/50 beam splitter is applied (which gives rise to the final state of the $x_{1,2}$ oscillators out of the virtual $x_\pm$ ones. Following \cite{Kim02} we can conclude that the non-classicality at the output modes (after the second beam splitter) must arise from some form of non-classicality at the input. This can exist if the equilibrium state has some degree of squeezing (this is the case for position coupling) or if the initial state is non-classical (either entangled or squeezed). The (pure) initial state which is least favorable for producing entanglement at the output are coherent states. The condition for the existence of entanglement in the final state for such initial states is $r_{crit}>1/2 \ln(2 \Delta x_+ \Delta p_+)$. Thus, to fulfill this condition we  need the environment to produce an equilibrium state where the variance of one of its quadratures is smaller than the vacuum limit,  i.e. $\min\{\Delta^2 x_+ , \Delta^2 p_+\}<1/2$  (for $m=1,\Omega_-=1$). We showed that this happens for position coupling and temperatures below $T_0$. Moreover, the above description of the problem enables us to draw a stronger conclusion: For symmetric coupling initial coherent states will never get entangled (even at intermediate times). This result, confirmed by our numerical simulations, 
can be seen as follows: For initial coherent states, the oscillators $x_+$ and
$x_-$ are not squeezed initially. Moreover, $x_-$ does not evolve for this type of coupling while the oscillator $x_+$ will change its variances but will never become squeezed due to the nature of the interaction, which is symmetric in position and momentum. Therefore, the two modes will have vanishing squeezing during the entire evolution and, as a consequence the oscillators $x_{1,2}$ will never be entangled. 

Finally, we studied the behavior of non-resonant oscillators by using both numerical and analytical tools. In this context we obtained a new master equation where the two virtual oscillators ($x_+$ and $x_-$) are coupled. Both oscillators approach an equilibrium state where they are not only correlated but also may have slightly different variances. 
We showed the existence of an entangled state at very low temperatures. Thus, we conclude that the generic fate of entanglement in a finite temperature environment is not only to become independent of the initial state. Also, we showed that there is a low temperature threshold that depends on the detuning above which the entanglement undergoes a sudden death. This is probably related with the entanglement studied in harmonic chains \cite{Anders08}.
A more detailed analysis of the possible scenarios for non-resonant oscillators will be presented elsewhere. 

\begin{acknowledgements}
JPP is a member of CONICET and AR acknowledge support from CONICET. This work was supported with grants from ANPCyT (Argentina) and Santa Fe Institute (SFI, USA).
\end{acknowledgements}

\end{document}